\documentclass[prd,nofootinbib,aps,floats,floatfix,amsmath,amssymb,secnumarabic,twocolumn]{revtex4} %
\usepackage{bm}
\usepackage{graphicx}
\usepackage[utf8]{inputenc}
\usepackage{color}
\usepackage{hyperref}
\usepackage{cancel}
\usepackage{verbatim}
\usepackage{epigraph}
\usepackage{slashed}
\newcommand{\be}{\begin{equation}}
\newcommand{\ee}{\end{equation}}
\newcommand{\bea}{\begin{eqnarray}}
\newcommand{\eea}{\end{eqnarray}}
\newcommand{\sss}{\scriptscriptstyle}
\newcommand{\sba}{s_{\sss\beta\alpha}}
\newcommand{\cba}{c_{\sss\beta\alpha}}
 
\newcommand{\cw}{c_{\sss W}}
\newcommand{\sw}{s_{\sss W}}
\newcommand{\phm}{\phantom{-}}
\newcommand{\rhoett}{\rho_e^{\tau\tau}}
\newcommand{\rhoetts}{\rho_e^{\tau\tau *}}
\newcommand{\ttpm}{\tau^+\tau^-}

\def\sfrac#1#2{{\textstyle{#1\over #2}}}

\def\nn{\nonumber}
\mathchardef\mhyphen="2D 
\newcommand\xhyphen{\mathop{\mhyphen}}

\begin{document}
\title{Scalar doublet models confront $\tau$ and $b$ anomalies}
\author{James M.\ Cline\footnote{jcline@physics.mcgill.ca}}
\affiliation{Department of Physics, McGill University,
3600 Rue University, Montr\'eal, Qu\'ebec, Canada H3A 2T8}
\affiliation{Niels Bohr International Academy \& Discovery Center, 
Niels Bohr Institute, University of Copenhagen, 
Blegdamsvej 17, DK-2100, Copenhagen, Denmark}

\begin{abstract}
There are indications of a possible breakdown of the standard model, suggesting
that $\tau$ lepton interactions violate flavor universality, particularly
through $B$ meson decays.  BABAR, Belle and
LHCb report high ratios of $B\to D^{(*)}\tau\nu$. There are long-standing
excesses in $B\to\tau\nu$ and $W\to\tau\nu$ decays, and a deficit in inclusive
$\tau$ to strange decays. We investigate whether two Higgs doublet models with
the most general allowed couplings to quarks, and a large coupling to $\tau$
leptons, can explain these anomalies while respecting other flavor constraints
and technical naturalness. Fits to $B\to D^{(*)}\tau\nu$ data require couplings
of the new Higgs doublet to down-type quarks, opening the door to many highly
constrained flavor-changing neutral current (FCNC) processes. We confront these
challenges by introducing a novel ansatz that relates the new up- and down-type
Yukawa couplings, and demonstrate viable values of the couplings that are free
from fine tuning. LEP and LHC searches for new Higgs bosons decaying via $H^0\to
\tau^+\tau^-$ and $H^\pm\to\tau^\pm\nu$  allow a window of masses $m_H =
[100$-$125]\,$GeV and $m_\pm\sim 100\,$GeV that is consistent with the
predictions of our model. Contamination of the $W^+\to\tau^+\nu$ 
signal by
$H^+\to\tau^+\nu$ decays at LEP could explain the apparent $W\to\tau\nu$ excess. We
predict that the branching ratio for $B_s\to \tau^+\tau^-$ is not far below its
current limit of several percent. An alternative model with decays of $B\to
D^{(*)}\tau\nu_s$ to a sterile neutrino is also argued to be viable.

\end{abstract}
\maketitle


\section{Introduction}

The origin of the Standard Model (SM) flavor structure is a mystery,
and any model predicting new patterns of flavor  violation must
confront very strong experimental bounds. This has given rise to the
Minimal Flavor Violation (MFV) paradigm
\cite{Chivukula:1987py,D'Ambrosio:2002ex,Cirigliano:2005ck,Kagan:2009bn}
as a guide for constructing new physics beyond the SM, that has been
highly influential in recent years.   MFV is extremely effective for
suppressing flavor changing neutral currents (FCNCs).  In this work we
confront some hints of new physics for which MFV seems generally too
strong to accommodate the observed deviations.  We are thus motivated
to consider an alternative that can allow for larger nonstandard
flavor effects.

Several recent experiments indicate possible 
deviations from the SM in some flavor-specific observables involving
$\tau$ leptons.
BaBar, Belle, and LHCb report the ratios $R(D)$ and $R(D^*)$, defined as
\bea
R(X) = \frac{\mathcal{B}(\bar{B} \rightarrow X \, \tau \, \bar{\nu})}{\mathcal{B}(\bar{B} \rightarrow X \, \ell \, \bar{\nu})}
\eea
where $\ell = e, \mu$. The summary of the SM predictions and the 
measurements is shown in table \ref{tab:anomalies}.
The reported measurements are consistent with each other, and
with previously reported results
\cite{Lees:2012xj,Huschle:2015rga,Aaij:2015yra}. The measurements are
also consistent with $e/\mu$ universality. However, the naively combined
experimental value for the ratio
$R(D^{(\star)})$ differs from the SM prediction
by more than $3\,\sigma$.

\begin{center}
\begin{table}[b]
\centering
\tabcolsep 8pt
\begin{tabular}{|c|c|c|}
\hline
- & $R(D)$ & $R(D^*)$ \\ \hline
{\rm SM} &  0.297 $\pm$ 0.017 & {0.252 $\pm$ 0.005}  \\
\hline
{\rm Belle} \cite{Huschle:2015rga} & 0.375 ${\scriptstyle\pm 0.064\pm
0.026}$ & 0.293 ${\scriptstyle \pm 0.038 \pm
0.015}$  \\
{\rm BaBar} \cite{Lees:2012xj} & 0.440 ${\scriptstyle\pm 0.058\pm
0.042}$
 & 0.332 ${\scriptstyle\pm  0.024 \pm
0.018}$ \\
{\rm LHCb} \cite{Aaij:2015yra} &  & 0.336 ${\scriptstyle\pm 0.027 \pm
0.030}$  \\
\hline
Expt.\  avg.: & 0.408 $\pm$ 0.050 & {0.321 $\pm$ 0.021} \\
\hline
\end{tabular}
\caption{Summary of experimental and predicted values for $R(D)$ 
and $R(D^*)$.}
\label{tab:anomalies}
\end{table}
\end{center} 

There have been other hints of a breakdown of lepton 
flavor universality between $\tau$ and $e/\mu$.  
The measured decay rate of $B\to \tau \nu$ displays some
tension with the SM prediction.  Although a recent measurement by 
Belle \cite{Adachi:2012mm} has reduced the discrepancy to the
level of $1.7\,\sigma$, the current world average measurement
remains a factor of 1.5 higher than the SM prediction
(see ref.\ \cite{Soffer:2014kxa} for a recent review.)
The observed rate of $W\to\tau\nu$ is  also in tension with
the standard model predictions: the LEP measurement is $\sim 10\%$
above the SM value, at $2.4\,\sigma$ significance.   The inclusive
decays of $\tau$ to strange quarks yield a value of the CKM matrix
element $V_{us}$ significantly lower than that required for
unitarity \cite{Lusiani:2015eja}.

A number of authors have studied $B\to D^{(*)}\tau\nu$ in the context
of type-III two Higgs doublet models (2HDMs), in which the most general couplings of fermions to
both doublets are allowed, as well as model-independent analyses that
include this framework
\cite{Fajfer:2012jt,Datta:2012qk,Crivellin:2012ye,Celis:2012dk,Tanaka:2012nw,
Freytsis:2015qca,Crivellin:2015hha}.  There are two possible operators contributing to
the hadronic part of these processes, mediated by charged Higgs
exchange, proportional to $C^{cb}_{S_R}\bar c_L\, b_R$ and  
$C^{cb}_{S_L}\bar c_R\, b_L$, respectively.  (For simplicity we assume
that the coefficients are real in the present work.)
Some studies  
\cite{Fajfer:2012jt,Crivellin:2012ye,Celis:2012dk,Tanaka:2012nw,
Crivellin:2015hha} 
found that $C^{cb}_{S_L}$ by itself is
sufficient to get a good fit to the observed decay rates.  However
several recent analyses \cite{Lees:2013uzd,Freytsis:2015qca} obtain
best-fit regions requiring $C^{cb}_{S_R}\sim - C^{cb}_{S_L}$.
In particular, these studies use not only the
total rates but also the differential decay distributions as inputs
to their fits, finding that $C^{cb}_{S_L}$ by itself does not fit the 
decay spectra.

This difference is crucial for model building, since having 
$C^{cb}_{S_L}\neq 0$ only requires that the new up-type Yukawa
matrix $\rho_u$ (which couples mainly to the nonstandard Higgs doublet) 
is important, while keeping the down-type couplings $\rho_d
\cong 0$.  If $C^{cb}_{S_R}$ is also large, then  $\rho_d\sim \rho_u$,
making it much more challenging to satisfy constraints on FCNCs. The
purpose of this paper is to see how far one can go toward overcoming
these challenges, within the context of 2HDMs, if the indication for $C^{cb}_{S_R}\sim -
C^{cb}_{S_L}$ persists in future analyses.

We will show that some of the flavor challenges can be addressed if
$\rho_u$ and $\rho_d$ are related to each other in a particular way
that involves the CKM matrix.  This is a new
ansatz for helping to give flavor protection to type III 2HDMs, 
which might be of interest more generally than for the particular
applications that motivated us here. It is quite different from
{MFV}, yet it appears to facilitate adequate control over FCNCs to make
the theory viable, especially in the
down-quark sector where the constraints are strongest.

The model is strongly constrained by LEP and LHC searches for the new
charged Higgs decaying into $\tau\nu$ and the neutral one decaying
to $\tau^+\tau^-$.   We find a window 
$\sim [100$-$125]\,$GeV of allowed masses for the new scalars that
passes the collider constraints while allowing for an explanation of the
$B$ decay anomalies.  Scalars of these masses are just beyond the 
kinematic reach of LEP, while being in a region of low efficiency for
LHC searches, if their couplings to quarks are sufficiently small.

The outline of the paper is as follows. In section \ref{model} we
define the model.  In section \ref{anomalies} we derive constraints on
the new Yukawa couplings $\rho_\nu$, $\rho_u$, $\rho_d$ arising from
fits to $R(D)$ and $B\to\tau\nu$, and a few key flavor-sensitive decays.
Section \ref{collider} examines the collider constraints determining the
allowed mass range of the new  $\sim 100\,$GeV  Higgs bosons.   
In section \ref{new} we present a novel ansatz relating $\rho_u$ and
$\rho_d$, that allows these constraints to be satisfied in a
controlled way.  It is  a linear relation involving the CKM matrix
$V$, a diagonal unitary matrix $U$, and an $O(1)$ parameter
$\eta$: $\rho_u^\dagger V = \eta\,U V\rho_d$. 

In section \ref{constraints} we calculate observables from  meson
oscillations that most strongly constrain the scenario, while in
section \ref{decays} we show that rare decay processes that might
challenge it are within the experimental limits.  Section
\ref{numerical} obtains a numerical fit to the couplings
$\rho_d^{ij}$, that determine $\rho_u$ through our ansatz. 
In section \ref{naturalness} we estimate the size of loop
contributions to the nonstandard Yukawa and Higgs couplings to establish
technical naturalness of the model.
In section \ref{uvmodel} we
outline a microscopic model that naturally implements the ``charge
transformation'' mechanism for relating $\rho_u$ and $\rho_d$ in the
manner of our ansatz. 
We outline an alternative version of the 
model in section \ref{neutrino}, where the leptonic coupling
$\rho_e$ is replaced by a coupling $\rho_\nu$ to neutrinos, assuming a 
light sterile neutrino
in the anomalous decays of $B$, rather than $\nu_\tau$.  This model
is less constrained by LHC searches for the neutral Higgs. 
Conclusions are given in section
\ref{conclusion}.  Details of the sterile neutrino version of the model
are given in the appendix.

\section{The model}\label{model}

We begin with the most general two Higgs doublet model, where
$H_1$ and $H_2$ are the doublets, each coupling to all the
fermions.  They have the conventional decomposition
\bea
H_1 = \frac{1}{\sqrt{2}} \, \left(\begin{array}{c} {\sqrt{2} \, H_1^+} \\
{v+ H_1^r + i H_1^i} \end{array} \right),  \quad
H_2 = \frac{1}{\sqrt{2}} \, \left(\begin{array}{c} {\sqrt{2} \, H_2^+} \\
{H_2^r + i H_2^i} \end{array} \right)
\eea
in terms of the real and imaginary parts of the neutral components.
The Yukawa coupling Lagrangian is
\bea\label{general}
{\cal L}_Y &=& -  { \bar Q}_{L} \, \hat y_u \, \tilde H_1 \, {u}_{R} -  
{ \bar Q}_{L} \, \hat y_d \,H_1 \, {d}_{R}  -  { \bar L}_{L} \, \hat y_e \,H_1 \,  {e}_{R},  \\
&&-  { \bar Q}_{L} \,  \hat\rho_u \,\tilde H_2 \, {u}_{R} -  
{ \bar Q}_{L} \, \hat\rho_d \,H_2 \, {d}_{R}  -  
{ \bar L}_{L} \,\hat\rho_e \, H_2 \,  {e}_{R} + {\rm h.c.} \nn
\eea
where flavor, color and ${\rm SU(2)}_L$ indices have been suppressed, and 
$\tilde H_i^a = \epsilon_{ab} H_i^{b\star}$.
The scalar Lagrangian is
given by
\bea
{\cal L}_S &=& |D^\mu H_1|^2 + 
|D^\mu H_2|^2- V(H_1,H_2)
\eea
where the potential is defined as
\bea\label{pot}
V &=& \lambda \, \left(H_1^{\dagger} \, H_1 - \frac{v^2}{2} \right)^2 + m_2^2 \, (H_2^{\dagger} \, H_2),  \\
&+& (m_{12}^2 \, (H_1^{\dagger} H_2) + {\rm h.c.}) + \lambda_1 \, 
(H_1^{\dagger} H_1) \, (H_2^{\dagger} H_2) \nn \\
&+& \lambda_2 \, (H_1^{\dagger}  H_2) \, (H_2^{\dagger}  H_1)  +
\left[\lambda_3 (H_1^{\dagger}  H_2)^2  + {\rm h.c.} \right] \nn \\
&+& \left[\lambda_4 (H_1^{\dagger} H_2) (H_2^{\dagger}H_2) 
+ \lambda_5 (H_2^{\dagger}H_1)(H_1^\dagger H_1) + {\rm h.c.} \right] \nn \\
&+& \lambda_6 (H_2^{\dagger} H_2)^2.\nn
\eea
In this basis of fields, $H_2$ has no vacuum expectation value,
requiring the condition $m_{12}^2+\lambda_5^\star v^2 /2=0$.  This
is just a choice of field coordinates, which in
general can always be achieved by doing a rotation
(conventionally denoted by angle $\beta$ as well as a possible rephasing) 
between $H_1$ and $H_2$; however in
section \ref{uvmodel} we will argue that the Yukawa couplings were
generated directly in this basis, so that
the $\hat y$ and $\hat\rho$ matrices can naturally have very different
magnitudes and structures.

For simplicity  we will assume that the potential (\ref{pot}) is
CP-conserving, so that there is no mixing between scalars and the
pseudoscalar.  
The rotation between the Higgs basis fields and the CP-even mass
eigenstates is
\bea \label{angledefn}
\left(\begin{array}{c} {H_2^r} \\
{H_1^r} \end{array} \right) = \left(\begin{array}{cc}  \cba & -\sba \\
 \sba & \phm\cba 
  \end{array} \right)  \, \left(\begin{array}{c} {h} \\
{H} \end{array} \right)
\eea
Here we have used notation that is conventional in 2HDMs, such that
for $\sba\cong 1$,  the SM-like Higgs boson $h$ is mostly
$H_1^r$.  
For small $|\cba|$, the mixing angle is approximately determined by
\bea
	\cba \cong {\lambda_5\,v^2\over 2(m_h^2-m_H^2)}
\label{mixing}
\eea
where the SM-like, new neutral and charged Higgs boson masses are 
respectively
\bea
	m_h^2 &\cong& 2\lambda v^2\nn\\
	m_H^2 &\cong& m_2^2 
	+\sfrac12 (\lambda_1 + \lambda_2 + 2\lambda_3)\,v^2
	\nonumber\\
	m_A^2 &\cong& m_H^2 - 2\lambda_3 v^2\nn\\
	m_\pm^2 &=& m_2^2 + \sfrac12  \lambda_1 v^2
\eea
These approximations are valid for small mixing.
We will also require that 
the splittings between masses of the neutral
scalars $H$, $A$ (CP-even and CP-odd respectively)
are small, so that they can be regarded as components of a
complex neutral field for most purposes.  
This not only simplifies the model but 
also proves 
useful for suppressing some
FCNC effects as we will show.
Small splittings are consistent with
$|\cba|\ll 1$ and $|\lambda_{3}| \ll 1$, since it can be shown 
(without any approximation) that
\bea
	m^2_H - m^2_A &=& \cba^2 (m_H^2 - m_h^2) + 2\lambda_3 v^2
\label{massrels}
\eea
(see for example \cite{Sierra:2014nqa}).
We will therefore assume that $|\lambda_3|\ll 1$ in addition
to $|\cba|\ll 1$.  
Although $\lambda_2$ could {\it a priori} be relatively large, 
in this work we will be interested in masses of order $m_H \lesssim m_h$
and $m_\pm \sim 100\,$GeV, corresponding to $\lambda_2\lesssim 0.2$.
 Electroweak
precision data (see eq.\ (10.26) of ref.\ \cite{PDG})  would allow for larger splittings, with $m_H$
as large as $175\,$GeV for $m_\pm \sim 100\,$GeV. 
The couplings 
$\lambda_{4},\,\lambda_6$ play no direct role for our predictions, but
can be relevant for understanding the expected size of radiative corrections
to the $\lambda_i$ couplings, as we will discuss in section
\ref{naturalness}.  Vacuum stability requires that 
$\sqrt{\lambda\lambda_6/2} > -\lambda_2$ if $\lambda_2<0$.

As usual, biunitary rotations on the quark fields in ${\cal L}_Y$ 
diagonalize $\hat y_u$, $\hat y_d$, $\hat y_e$, with the scalar doublets still in the Higgs basis.
The subsequent rotation (\ref{angledefn}) then brings ${\cal L}_Y$ 
to the form 
\bea
	{\cal L}_Y &=& -{1\over\sqrt{2}}\sum_{\phi=h,H,A\atop f=u,d,e} y^f_{\phi ij}\, 
	\bar f_i \phi  	P_{\sss R} f_j +{\rm h.c.} \\
	y^f_{h ij} &=& \sba {\sqrt{2} \, m_f^i\over v}\delta_{ij} + \cba \rho_f^{ij},  \\
	y^f_{Hij} &=& \cba {\sqrt{2} \, m_f^i\over v}\delta_{ij}
	-\sba \rho_f^{ij}, \\
	y^f_{Aij} &=& \rho_f^{ij}\times \left\{{\textstyle{\!\!\!+i,\  f=u\atop -i,\  f=d,e}}
	\right.\\
 	&\phantom{=}&\nonumber
\eea 
where $P_{\sss R} = (1+ \gamma_5)/2$ is the usual chiral projector 
and $v \simeq 246\,$GeV (see for example the discussion in
\cite{Omura:2015nja}). 
The matrices $\rho_f^{ij}$ with $f=e,u,d$ are in general complex and
can induce tree-level FCNCs. 
They are given explicitly by 
\bea
\rho_u &=& L_u^\dagger \, \hat\rho_u \,  R_u, \nn \\
\rho_d &=& L_d^\dagger \, \hat\rho_d \,  R_d, \nn \\
\rho_e &=& L_e^\dagger \, \hat\rho_e \,  R_e,\nn
\eea
where the unitary matrices transform between the weak and mass 
eigenstates, and determine the CKM matrix $V = L_u^\dagger L_d$.
The charged scalars couple to the fermions as 
\bea
\label{chargedL}
	{\cal L} &=& -\bar\nu\left(U_{\nu}^\dagger\,\rho_e\right) H^+\, P_{\sss R}\, e\\
	&\, & - \bar u\left( V\rho_d P_{\sss R}
	- \rho_u^\dagger VP_{\sss L}\right) H^+ d
	+{\rm h.c.}\nonumber
\eea
where $U_{\nu} = L_\nu^\dagger \, L_e$ is the 
PMNS neutrino mixing matrix.  Since neutrino oscillations are 
unimportant in 
the processes under consideration, we 
henceforth replace $U_{\nu}\,\nu \to \nu$ with the understanding that 
$\nu$ refers to the 
initially emitted flavor eigenstate.

\section{Explaining the anomalies}
\label{anomalies}

Our primary motivation is to present a framework that is able to
simultaneously explain the excess signals in processes with
final state $\tau$ leptons: $B\to D^{(*)}\tau\nu$, $B\to\tau\nu$
and $W\to\tau\nu$.  In addition we consider the hint of a deficit
in $\tau\to K^-\nu$ decays. 
In this section we will show how these can come about
at tree level due to exchange (or decay) of the charged Higgs $H^\pm$,
for appropriate choices of the new Yukawa
couplings in $\rho_e$, $\rho_d$ and $\rho_u$.  The decays
of $B$, $B_s$ and $h$ into $\tau^+\tau^-$ provide immediate constraints on 
the scenario, which we therefore also consider in this section.

\subsection{$B\to D \tau\nu$, $B\to D^{(\star)} \tau\nu$}
New contributions to $B\to D \tau\nu$ can be mediated by the
tree-level exchange of the charged Higgs $H^\pm$ if $\rho_e^{i\tau}$
is nonzero, as can be seen from eq.\ (\ref{chargedL}).  The matrix element $\rhoett$ turns out to be the optimal
choice for satisfying the combined constraints from LHC searches
for the neutral boson $H^0$ and rare leptonic decays of $B$ and
$B_s$ mesons.  We will therefore assume that 
$\rhoett\neq 0$, while 
the remaining entries in $\rho_e^{ij}$ are very small or vanishing.
 
Integrating out the $H^\pm$ 
then produces the effective Hamiltonian
\bea
	\mathcal{H} &=& {\rhoetts\over m_\pm^2}\, 
	\left[\bar \tau 
	P_{\sss L}\nu_\tau
\right] \left[\bar c\left(V\rho_dP_{\sss R} - 
	\rho_u^\dagger V P_{\sss L}\right)^{cb} b\right]\nonumber\\
	&\equiv& 
	{1\over \Lambda^2}(\bar\tau P_{\sss L}\nu_\tau)
	\left[\, C^{cb}_{S_R}(\bar c P_{\sss R} b)
	 + C^{cb}_{S_L}(\bar c P_{\sss L} b) \,\right]
\label{BDham}
\eea
that is relevant for $b\to c\tau\nu$ at the quark level.
Ref.\ \cite{Freytsis:2015qca} performed a 
fit to the $B\to D^{(*)} \tau\nu$
rates  and decay spectral using the two operators in (\ref{BDham}),
which interfere with the standard model contributions.
Two viable solutions for the Wilson coefficients were found 
there, of which the smaller ones correspond to
\bea
\label{BDfits2a}
	{C^{cb}_{S_R}\over\Lambda^2} &\cong& {(\rhoett)^* 
	(V\rho_d)^{cb}
	\over m_\pm^2}\ \ \,\cong\ \  {1.25\over {\rm TeV}^2}\\
\label{BDfits2b}
	{C^{cb}_{S_L}\over\Lambda^2} &\cong& -{(\rhoett)^* 
	(\rho_u^\dagger V)^{cb}
	\over m_\pm^2}
	\cong -{1.02\over {\rm TeV}^2}
\eea
There is an intriguing relationship between the 
couplings,
\be
	(V\rho_d)^{cb} \cong (\rho_u^\dagger V)^{cb}
\label{hint1}
\ee
about which we will say more below.

\subsection{$B\to \tau\nu$}

The contribution of the new charged Higgs to 
$B^+\to \tau^+\nu$ decay	
modifies the branching ratio (BR) as  
\cite{Hou:1992sy,Fajfer:2012vx,Crivellin:2012ye}
\bea
\mathcal{B}(B^+ \rightarrow \tau^+\nu) &=& 
\frac{G_F^2 |V_{ub}|^2}{8 \, \pi} m_\tau\, \tau_B\, f_B\, m_B \,
\left(1- \frac{m_\tau^2}{m_B^2}\right)^2 \, \nn \\
&\times& \left|1+ \frac{m_B^2}{\bar m_b \, m_\tau} 
\frac{(C^{ub}_{S_R} - C^{ub}_{S_L})}{C^{ub}_{SM}}\right|^2
\label{Btaunu}
\eea  
where $\bar m_b$ is the $b$-quark $\overline{\rm MS}$  mass,
$C^{ub}_{S_R}$ and $C^{ub}_{S_L}$ are defined analogously
to $C^{cb}_{S_R}$ and $C^{cb}_{S_L}$ in eq.\ (\ref{BDham})
 and $C^{ub}_{SM} = 4 G_F \, V_{ub}/\sqrt{2}$. 

To estimate the possible allowable size of the new physics (NP)
contribution, we take the enhancement factor in the second line of 
(\ref{Btaunu}) to be less than 2.6, the ratio between the 
$3\sigma$ maximum allowed value of
the world average measurement $(1.14\pm0.27)\times 10^{-4}$ and the 
CKMfitter prediction \cite{Charles:2015gya} $0.76\times 10^{-4}$ of 
the BR.  This gives the bounds
\be
	-{0.02\over
	{\rm TeV}^2}  \lesssim   {\rhoett\over m_\pm^2}\left(
     (V\rho_d)^{ub} + (\rho_u^\dagger V)^{ub}\right)
      \lesssim {0.05\over
	{\rm TeV}^2}
\label{Btaunu_bound}
\ee
Curiously, this suggests a relation similar to (\ref{hint1}), but with the
opposite sign,
\be
	(V\rho_d)^{ub} \cong -(\rho_u^\dagger V)^{ub}
\label{hint2}
\ee
In section \ref{new} we will present an ansatz that combines these
two conditions in a concise way.

If the indication for a 
factor of 1.5 excess in the observed versus SM predicted partial width
is interpeted as evidence for new physics, then the 
condition (\ref{hint2}) is not a strict equality, and 
we should replace (\ref{Btaunu_bound}) with the condition
\be
\left|{(V\rho_d)^{ub} + (\rho_u^\dagger V)^{ub}
\over (V\rho_d)^{cb} + (\rho_u^\dagger V)^{cb}}\right|
      \cong 8\times 10^{-3}
\label{Btaunu_target}
\ee
where we used (\ref{BDfits2a}-\ref{BDfits2b}) to eliminate $\rhoett/m_\pm^2$.

\subsection{$W\to\tau\nu$}
\label{Wtaunu}

The branching ratios for $W\to\ell\nu$ were measured by the LEP experiments 
for the individual lepton flavors, from the production of $W^+ W^-$ pairs.  
The averaged results are \cite{PDG}
\bea
B(W\to e\nu) &=&  (10.71\pm 0.16)\%\nn\\
B(W\to \mu\nu) &=&  (10.63\pm 0.15)\%\nn\\
B(W\to \tau\nu) &=&  (11.38\pm 0.21)\%
\eea
The ratio of decays to $\tau\nu$ versus the first two generations is
\be
	R_{\tau/\ell}= {B(W\to\tau\nu)\over\sfrac12 [ B(W\to e\nu)+B(W\to\mu\nu)]}
	= 1.066\pm 0.028
\ee
which deviates from 1 by $2.35\,\sigma$.  It was suggested by refs.\ 
\cite{Dermisek:2008dq,Park:2006gk} that the excess could be due to contamination of the
$W$ decay signals by charged Higgs bosons with mass close to $m_W$ decaying to $\tau\nu$.
In a detailed reanalysis of data 
reported by DELPHI, it was found that the discrepancy could be reduced to
1.03 for a charged Higgs mass of $m_\pm=81\,$GeV and $B(H\to\tau\nu)=0.7$,
$B(H\to qq)=0.3$, values ruled out by the more recent LEP study 
\cite{Abbiendi:2013hk}.  Ref.\ \cite{Dermisek:2008dq} found that for
$B(H\to\tau\nu)=1$, which is the appropriate limit for our model, the
observed $R_{\tau/\ell}$ could be explained if $m_\pm = 94{+4\atop-3}\,$GeV
in the $2\sigma$ region.  This is marginally compatible with the combined
LEP limit of $m_\pm < 94\,$GeV \cite{Abbiendi:2013hk}.

More recent LHC measurements of $W\to\tau\nu$ \cite{Czyczula:2015mfa} do 
not see evidence for any excess, but this does not contradict having an
observable effect at LEP since the production mechanism for $H^\pm$ in this
case depends upon its coupling to quarks, which is very small in our model.
The same remark applies for the Tevatron, where no such effect was 
observed either. 

We do not attempt to reanalyze the LEP signal here, but point out one
potentially important difference between our model and those considered
previously in this context.  We will require a sizable coupling $\rhoett
\cong 2.5$ leading to a large width for the charged Higgs,
\be
	\Gamma_{H^\pm} = {|\rhoett|^2\over 16\pi}m_\pm
	\cong 12\,{\rm GeV}
\ee
for $m_\pm = 100\,$GeV.  This could allow for a greater effect on
$R_{\tau/\ell}$ with $m_\pm > 95\,$GeV than would be possible in the usually
assumed case where $H^\pm$ width effects are ignored.

\subsection{$\tau\to K^-\nu$}
\label{tauKnu}
Decays of $\tau$ into strange particles are among the processes used
to determine the CKM matrix element $V_{us}$.  The HFAG collaboration
recently 
noted that the inclusive decays of this kind lead to a determination
that is $3.4\,\sigma$ below the value consistent with unitarity, while the
average from all $\tau$ decays is $2.9\,\sigma$ too low 
\cite{Lusiani:2015eja}.  This discrepancy
could be explained if the contributions from charged Higgs exchange
interfere destructively with the SM amplitudes.

Focusing on the specific decay $\tau\to K^-\nu$, the new contribution
to the amplitude is given by 
\be
\label{MtauKnu}
	{\cal M}_{\tau\to K\nu} = 
	\,{\rhoett(V\rho_d +\rho_u^\dagger V)_{us}\, f_K\, m_K^2
	\over 2\, m_\pm^2\, (m_u+m_s)}\, 
	(\bar u_\tau P_{\sss L} u_\nu)
\ee

Taking the central values of $|V_{us}|=0.2211\pm 0.0020$ determined from
$\tau\to K\nu$ and $|V_{us}|=0.2255\pm 0.0010$ from CKM unitarity
\cite{Lusiani:2015eja}, we 
estimate that 
\be
	(V\rho_d+\rho_u^\dagger V)_{us}\cong -4.2\times 10^{-4}
\label{Vruseq}
\ee
(assuming fiducial values  $\rhoett=2.5$ and $m_\pm = 100\,$GeV
that will be preferred below).
The minus sign is necessary to get 
destructive interference between $W^\pm$ and $H^\pm$ exchange, given
that the relative signs of $\rhoett$ and $V\rho_d$ are fixed
by requiring constructive inteference in $B\to D^*\tau\nu$ decays.

We will find that there is some mild tension between (\ref{Vruseq})
and other observables (notably $b\to s\gamma$) in the numerical fit to be described in 
section \ref{numerical}, so that we do not insist on this potential
anomaly in our fits.  However it can plausibly fit
into the general pattern of deviations in $\tau$ interactions that
are addressed by our model.

\subsection{Constraint from $h\to\ttpm$}

Because of mixing in the scalar sector, the SM-like Higgs acquires
small additional couplings $\cba\rho_f$ to fermions.
 They contribute to the partial width of $h$ as 
\bea
\Gamma(h \rightarrow f_i \, \bar f_j + f_j\bar f_i) &=& N_c \, 
\frac{m_h}{16 \, \pi} \, \left(|\sba y_{ij}
+ \cba\rho_{ij}|^2\right.\nn\\ && \left.
	\qquad\qquad + \{i\leftrightarrow j\}\right)
\eea
for all kinematically accessible final states (with $N_c$
colors), where $y_{ij} = \sqrt{2} m_i\delta_{ij}/v$. 
In our model, $\rhoett$ is 
the largest new coupling.  

New contributions to the decays into $\tau^+\tau^-$ are 
constrained by ATLAS and CMS observations  \cite{Higgs_constraints}. Deviations from the
SM expectation are characterized by a coupling modifier $\kappa_\tau
= |1 + \cba\rhoett/y_\tau|
\in [0.64,\,1.14]$ at $2\sigma$, where $y_\tau$ is the SM Yukawa coupling.
We get the least restrictive constraint if
$\cba\rhoett$ is negative, in which case there are two solutions,
\bea
	|\cba\,\rhoett| &<& 
	3.7\times 10^{-3}\nn\\
	-0.022\  <\  	\cba\,\rhoett &<& -0.017
\label{rhoett_bound}
\eea
at 95\% confidence level.  Later we will adopt a fiducial value of 
$\rhoett=2.5$.  In that case the central value of
(\ref{rhoett_bound}) implies $\cba = -7.8\times 10^{-3}$.  This region
corresponds to the amplitude for $h\to\ttpm$ having the opposite sign
relative to the SM value.

\subsection{Constraints from $B,B_s,\Upsilon\to\ttpm$}

The considerations leading to (\ref{BDfits2a}) fix only 
a linear combination of the $\rho_d$ couplings, namely
$V_{cd} \rho_d^{db} + V_{cs} \rho_d^{sb} + V_{cb} \rho_d^{bb}$.
It is useful at this point to notice that $\rho_d^{db}$ is 
strongly constrained by the upper  limit of $4.1\times 10^{-3}$
\cite{PDG}
on the BR for $B^0\to\tau^+\tau^-$.  In our model,
this decay is mediated by $H^0$ exchange, with rate
\be
	\Gamma(B\to\ttpm) \cong {|\rhoett
	\rho_d^{db}|^2\,f_B^2\, m_B^2\,(m_B^2-2m_\tau^2)^{3/2}
	\over 64\pi\, \bar m_b^2\, m_H^4}
\label{bmutau}
\ee
where $f_B = 0.19{\rm\,GeV}$ is the $B$ decay constant 
\cite{Na:2012kp}.
If we tried to satisfy the constraint in
(\ref{BDfits2a}) with only $\rho_d^{db}$
nonvanishing, anticipating that $m_\pm\sim m_H\sim 100\,$GeV 
 (see section \ref{collider}), eq.\  
(\ref{bmutau}) would then imply that $\Gamma(B\to\ttpm)$ is as large
as the total measured width of $B^0$.  We find the upper limit
\be
	|\rho_d^{db}| < 1.0\times 10^{-3}\,\left(m_H\over
100{\rm\,GeV}\right)^2\left|2.5\over \rhoett\right|
\label{rhoddb_bound}
\ee

\begin{figure*}[t]
\centerline{
\includegraphics[width=0.9\columnwidth]{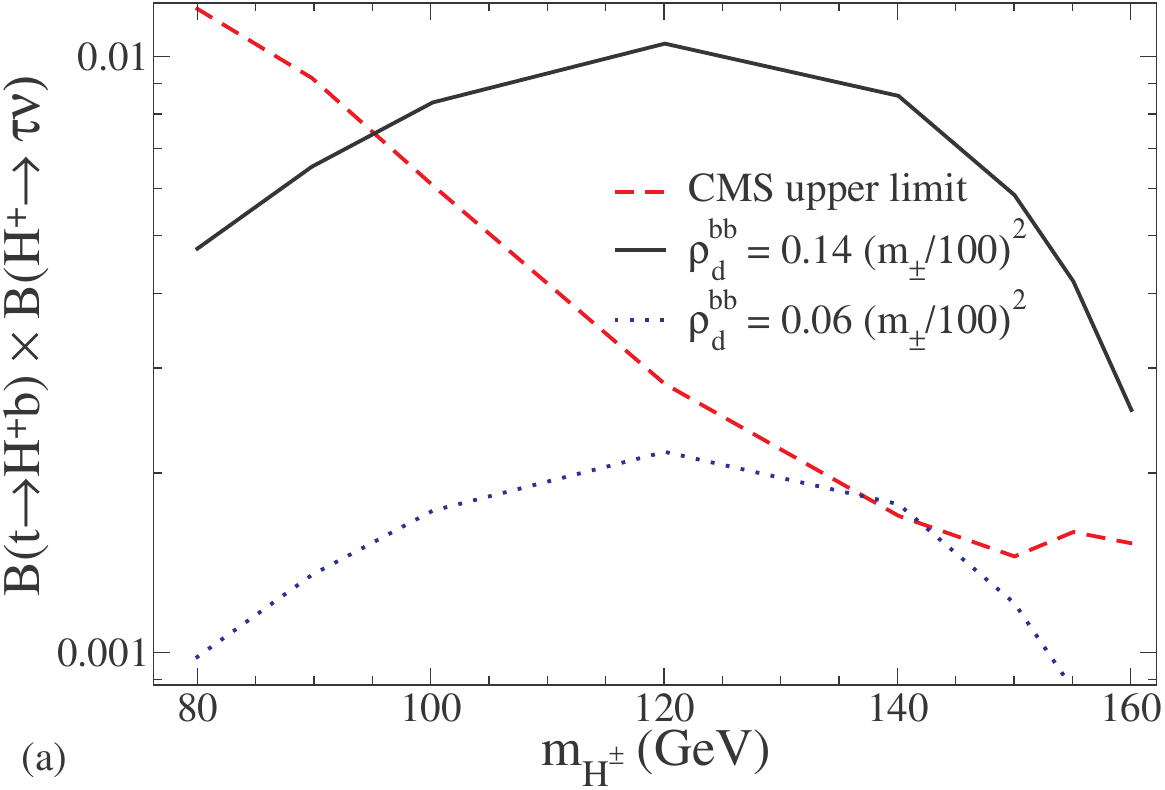}
\hfil
\includegraphics[width=0.9\columnwidth]{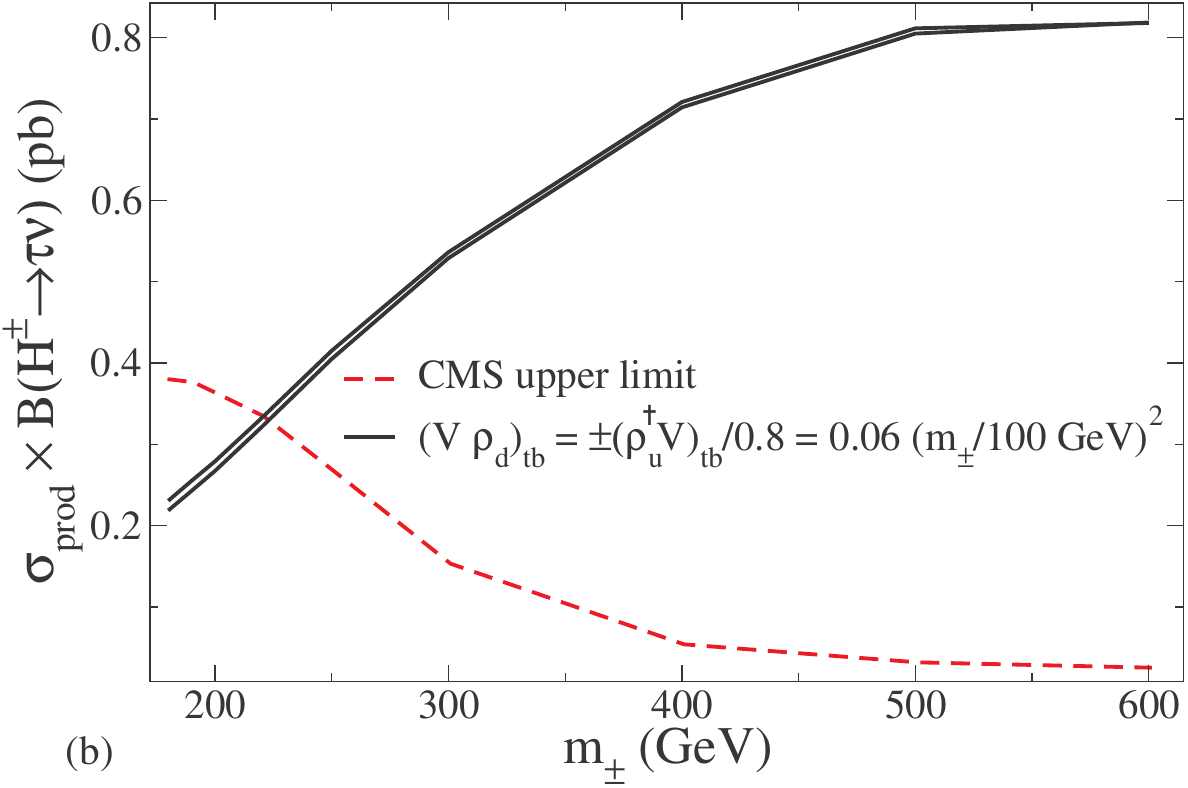}}
\caption{Left (a): CMS constraints \cite{Khachatryan:2015qxa} on a 
charged Higgs decaying to $\tau\nu$ in the $m_\pm <m_t$ and 
$m_\pm > m_t$ regions.
Left (a): upper limit on $B(t\to b H^+)\cdot B(H^+\to\tau^+\nu)$
versus $m_\pm$ in the low mass region.  Right (b): upper limit on 
production cross section times $B(H^+\to\tau^+\nu)$ for 
$m_\pm > 180\,$GeV.  Assumed values of the $\rho_d^{bb}$ coupling
from eq.\ (\ref{rhodbb_const}) are indicated.}
\label{Hpmlimit}
\end{figure*}

Similarly, trying to use  $\rho_d^{sb}$ to saturate (\ref{BDfits2a})
results in a branching ratio of $0.1$ for  $B_s\to\tau^+\tau^-$. 
However the current limits on this decay channel are very weak,
$B(B_s\to\tau^+\tau^-)< 5\%$ \cite{Grossman:1996qj,Blake:2015tda}.
Using $f_{B_s}=0.225\,$GeV, this gives 
a bound on $|\rho_d^{sb}|$ of 
\be
	|\rho_d^{sb}| < 2.8\times 10^{-3} 
\left(m_H\over
100{\rm\,GeV}\right)^2\left|2.5\over \rhoett\right|
\left(B(B_s\to\ttpm)\over 5\times 10^{-2}\right)^{1/2} 
\label{rhodsb_bound}
\ee
It follows that we must
rely upon $\rho_d^{bb}$ to provide at least part of the
contribution to the Wilson coefficient $C_{S_R}$, if we
insist on its central value from (\ref{BDfits2a}). 
This gives the constraint
\be
	|\rho_d^{bb}|\left(100{\rm\,GeV}\over m_\pm
\right)^2\left|\rhoett\over 2.5\right| \cong [0.05-0.12]
\label{rhodbb_const}
\ee
where the lowest value in the interval corresponds to saturating
the limit (\ref{rhodsb_bound}).

With $\rho_d^{bb}\neq 0,$ neutral $H^0$ exchange also leads to the decays of the $b\bar b$ bound
state $\chi_{b0}\to\ttpm$ at tree level.   In the SM such decays
are dominantly electromagnetic, which greatly suppresses the BR of
the $H^0$-mediated process. No bounds on leptonic decay modes of
$\chi_{b0}$ are given by the Particle Data Group.  For $\Upsilon$
the 
branching ratio for  $\ttpm$ final states is $(2.60\pm 0.10)\%$ 
but since it
is a vector, $H^0$ cannot mediate the decay at tree level.  Rather
it proceeds at one loop with virtual $H^0$ and photon exchange.
We estimate that it contributes less than $10^{-14}$ to the branching
ratio.

\begin{figure*}[t]
\centerline{
\includegraphics[width=0.85\columnwidth]{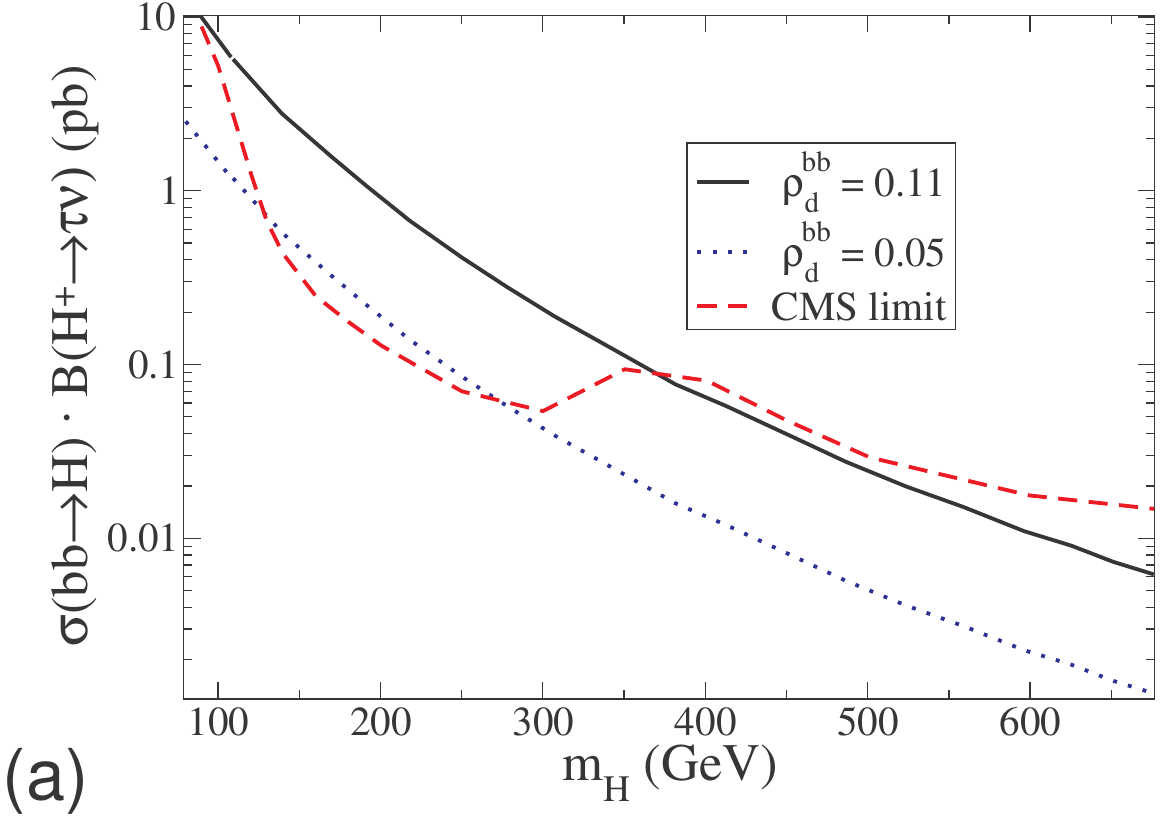}\hfill
\includegraphics[width=1.1\columnwidth]{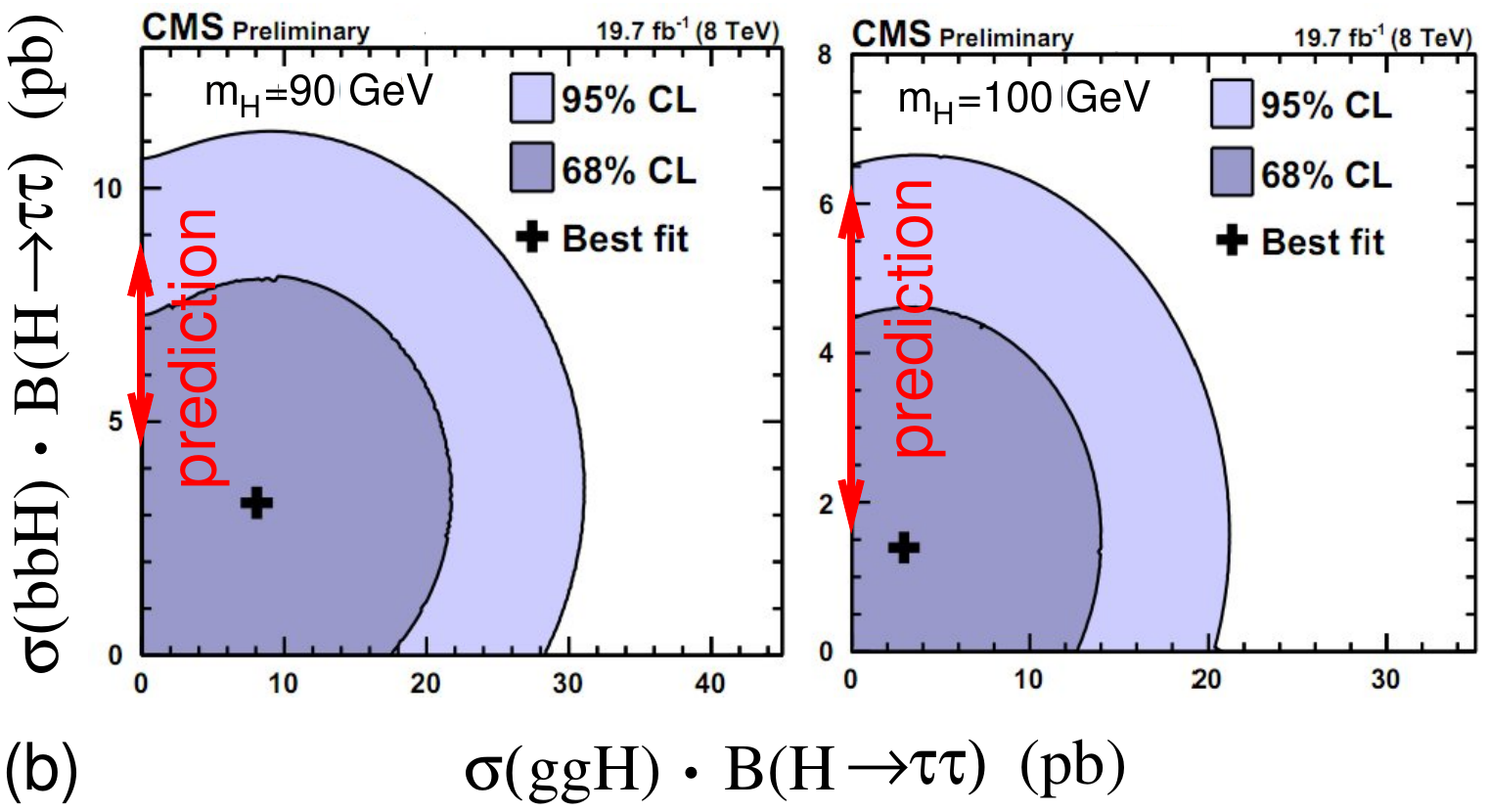}}
\caption{Left (a): predicted $bb$ fusion cross section versus neutral
Higgs mass for $\rho_d^{bb}=0.05$ and 0.11, and CMS preliminary 
upper limit
\cite{CMS:2015mca}. Right (b): allowed regions in the plane of $bb$
fusion versus $gg$ fusion cross sections, for $m_H = 90,100\,$GeV.
Theoretical predictions for $0.07\, (0.05) < \rho_d^{bb}<0.10$ are 
indicated for $m_H = 90\, (100)\,$GeV,
having negligible $\sigma(ggH)$.  Figure adapted from ref.\ \cite{CMS:2015mca}.}
\label{Htautau}
\end{figure*}

\section{Collider constraints}
\label{collider}

We next consider LEP and LHC searches for charged and neutral Higgs
bosons with decays principally into $\tau$ leptons, as predicted in our
model.  The charged Higgs can also have an indirect signature through its
effect on the $h\to \gamma\gamma$ partial width.

\subsection{Charged Higgs searches}

ATLAS \cite{Aad:2014kga} and CMS \cite{Khachatryan:2015qxa} have 
recently reported on searches for charged Higgs  particles decaying
into $\tau\nu$, which is  the principal decay channel of $H^\pm$
in our model.  These searches constrain our scenario in the
mass ranges $m_\pm \in [80,\,160]$ GeV and [180,\,1000] GeV,
complementing previous LEP studies that excluded $m_\pm < 95\,$GeV
\cite{{Abbiendi:2013hk}} for $H^+$ decaying with branching ratio
of 100\% into $\tau^+\tau^-$ as is the case in our model.

At low masses $m_\pm < 160\,$GeV, the
product of branching ratios $B(t\to H^+b)\cdot B(H^+ \to \tau^+\nu)$
is bounded, since the dominant production process is through
top quark decays into $H^\pm b$.  Our model predicts that
\bea
	B(t\to H^+ b) &\cong& {(1+\eta^2)(\rho_d^{bb})^2\over
	32\pi\,m_t^3\,\Gamma_t}(m_t^2-m_\pm^2)^2\nonumber\\
	B(H^+\to \tau\nu) &=& 
	\left(1 +3|\rho_d^{bb}/\rhoett|^2\right)^{-1}
	\cong 1
\label{BRs}
\eea 
ignoring $m_b$, where $\Gamma_t = 1.4\,$GeV is the measured width of
the top quark.\footnote{The NP contribution to $\Gamma_t$ must be less
than $1\%$ in the experimentally allowed region.}
The prediction is plotted along with the CMS limit in fig.\
\ref{Hpmlimit}(a), using the upper and lower values of
$\rho_d^{bb}$ consistent with $R(D)$ from eq.\ (\ref{rhodbb_const}).
For the smaller value of $\rho_d^{bb}$, there is almost
no restriction on the allowed mass $m_\pm$.  The D$\slashed{0}$ collaboration
finds a much weaker limit on 
$B(t\to H^+b)\cdot B(H^+ \to \tau^+\nu) \lesssim 0.2$ in this mass range
\cite{Abazov:2009aa}. 
CDF obtains the stronger limit of 0.06 \cite{Aaltonen:2014hua}, which however is still not
competitive, and we do not show the Tevatron limits on the plot.

At higher masses $m_\pm > 180\,$GeV, the $H^\pm$ is produced by
its coupling to $tb$, either through $gg\to H^+\bar t b$ or
$g\bar b\to H^+\bar t$.  In fig.\ \ref{Hpmlimit}(b)
we compare the CMS bound to the model predictions taking the lower value of $\rho_d^{bb}$ 
indicated in eq.\ (\ref{rhodbb_const})\footnote{We thank Grace Dupuis
for computing this production
cross section using MadGraph}\ \ A charged Higgs mass up to 220
GeV could be consistent with this search.

\subsection{Neutral Higgs searches}

ATLAS and CMS searches for the neutral $H^0$ decaying to 
$\tau^+\tau^-$ \cite{Khachatryan:2014wca, Aad:2014vgg,CMS:2015mca} put much
stronger constraints  on our model, forcing us to consider low values
of both $m_H$ and $m_\pm$.  Since the $\rho_d^{bb}$ coupling scales as
$m_\pm^2$ to fit $R(D)$ (eq.\ (\ref{rhodbb_const})), $H^0$ typically
has a larger coupling to $b$ quarks than does the SM Higgs boson. As
a result, neutral $H^0$ production by gluon-gluon fusion
\cite{Dawson:1990zj,deFlorian:2009hc,FGupdate}, which is the
dominant process for the SM Higgs, can be small compared to $bb$ fusion, 
leading to strong constraints on the $\sigma(bbH)$ cross section.
These limits are weakest at low $m_H$, and also at low $m_\pm$ due to
the scaling of $\rho_d^{bb}\propto m_\pm^2$. 

In fig.\ \ref{Htautau}(a) we plot the predictions of our model for 
$\sigma(bbH)$ versus $m_H$ (note that $B(H\to\ttpm)=1$ to a very
good approximation), using the values $\rho_d^{bb}=0.05$ and $0.11$
suggested by eq.\ (\ref{rhodbb_const}).
To compute $\sigma(bbH)$, we rescaled the cross sections obtained in
ref.\  \cite{Trott:2010iz} (which are computed for a range of $m_H$)
by the more accurate recent results (computed at a few values of
$m_H$)  in ref.\ \cite{Bonvini:2015pxa}.
Only for the lower value of $\rho_d^{bb}$ are there any regions
consistent with low $m_H$.  Large values of $m_H$ cannot be
reconciled with $\rho_d^{bb}$ as small as assumed ($\sim 0.1$) because
of the $\rho_d^{bb}\propto m_\pm^2$ scaling, and the need to keep
$|m_\pm - m_H|\lesssim 75\,$GeV to respect electroweak precision
constraints.   In the optimistic
case of $\rho_d^{bb} = 0.05$, we find an upper limit of $m_H < 125\,$
GeV.  

\begin{table}[b]
\begin{tabular}{|c||c|c|c|c|c|}
\hline
$m_H$ & $90.0$ & $92.5$ & $95.0$ & $97.5$ & $100.0$ \\
\hline
$S_{95}$   & 0.39 &   0.70 &  1.07 &  2.88 & 5.29 \\
\hline
\end{tabular}
\caption{LEP limits \cite{Schael:2006cr} on the production cross section of 
neutral Higgs bosons from
pair production $e^+e^-\to H H^*\to \ttpm\ttpm$, as a function of their mass 
$m_H$ in GeV.  $S_{95}$ is the 95\% c.l.\ upper limit on the ratio of
the observed cross section to the predicted one.}
\label{tab:lep}
\end{table}

The CMS search has marginal evidence for excess events at $m_H \cong
90-100\,$GeV, as shown in fig.\ \ref{Htautau}(b).  There is a
slight preference for nonzero values of the two production cross
sections $\sigma(ggH)$ and $\sigma(bbH)$.  Our model predicts very
small values of the former, $\sim 0.1\,$pb, but significant values
of $\sigma(bbH)$.  We show the range of predictions corresponding to
$\rho_d^{bb} = 0.05$ to $0.10$ by the vertical arrows.  The lower value corresponds to 
saturating the limit on $B_s\to\ttpm$ in eq.\ (\ref{rhodsb_bound})
in order to make $\rho_d^{bb}$ as small as possible in eq.\ 
(\ref{rhodbb_const}).  The higher value corresponds to a branching
ratio for $B_s\to\ttpm$ of $5\%/\sqrt{2} = 3.5\%$.  There is
a strong correlation between $B(B_s\to\ttpm)$ and the possibility
to satisfy the CMS constraint, leading to our prediction that
$B(B_s\to\ttpm)$ cannot be much smaller, unless the evidence for
$C_{S_R}$ from $R(D)$ (eq.\ (\ref{BDfits2a})) becomes
weaker.\footnote{At $m_H=125\,$GeV, with $\rhoett=2.5$, 
we can obtain $\rho_d^{bb} = 0.07$ with $B(B_s\to\ttpm) = 2\%$.
Lower values of $m_H=125$ require larger $B(B_s\to\ttpm)$.}
 
LEP also constrained the neutral Higgs boson mass in the case of
interest for our model, where $H\to\ttpm$ almost exclusively
\cite{Schael:2006cr}.  The statistic $S_{95}$ is defined as the
95\% c.l.\ upper bound on the production cross section of $H^0$
pairs, in units of the theoretical cross section for 
$e^+e^-\to Z^*\to H H^*$.  In the LEP analysis it was assumed that 
both neutral Higgs
bosons are nearly degenerate, which is the same assumption that we
make in our model.  The pair production cross section
 is model-independent, since it depends
only upon the SU(2) gauge interactions of the extra scalar doublet. 
$S_{95}$ versus $m_H$ is listed in table \ref{tab:lep}, showing that
$m_H$ must be greater than $95\,$GeV.  For $m_H\ge 100\,$GeV, the allowed
cross section is more than 5 times greater than predicted, allowing for overlap
between the LEP- and CMS-allowed regions.

CDF constrained the gluon fusion cross section $\sigma(ggH)$ times $B(H\to
\ttpm)$ to be less than $1.7\,$pb for $m_H = 115\,$GeV 
\cite{Aaltonen:2012jh}, while the SM prediction is $\sigma(ggH)_{\rm SM} = 
1.07\,$pb \cite{Baglio:2010um}.  In our model 
$\sigma(ggH)/\sigma(ggH)_{\rm SM}= (\rho_u^{tt}/y_t)^2 \cong 2\times
10^{-3}$, far below the CDF limit.  Similarly to LHC, searches for $H\to\ttpm$
in association with $b$ quarks are more sensitive.  D$\slashed{0}$ constrained
$\sigma(bbH)\cdot B(H\to\ttpm) < 0.8\,$pb for $m_H = 115\,$GeV 
\cite{Chakrabarti:2011rv}.  The SM cross section is $\sigma(bbH)\sim 6\,$fb
near this mass \cite{Dittmaier:2003ej}, which gets scaled by
$(\rho_d^{bb}/y_b)^2 \cong 5$ in our model, again much smaller than the
limit.

In summary, $m_\pm \cong 100\,$GeV and $m_H \cong [100$-$125]\,$GeV
are the favored mass ranges for satisfying the combined limits
from LEP and LHC, subject to the constraints from flavor physics
discussed in section \ref{anomalies}.  A large value
$\rhoett = 2.5$ is also needed, which we will show is
allowed by lepton flavor universality of $Z\to\ell\ell$ decays, 
in section \ref{Zll}.  Larger values of $m_\pm$ require larger
values of $\rho_d^{bb}$ to explain $R(D)$, making it more
difficult to respect searches for the neutral Higgs.

\subsection{Charged Higgs contribution to $h\to\gamma\gamma$}
The charged Higgs contributes to $h\to\gamma\gamma$ at one loop,
with an amplitude that is proportional to 
\bea
	A &\sim& 3 Q_t^2 A_{1/2}(\tau_t) + A_1(\tau_W) + g_\pm A_0(\tau_\pm)
	\nonumber\\
	A_0(\tau) &=& -\tau^{-2}\,(\tau - {\rm arcsin}(\sqrt{\tau})^2)
\eea
(see for example 
ref.\ \cite{Posch:2010hx})
where the first two terms are from the top quark and $W$ boson loop,
giving $-6.5$, while $\tau_\pm = (m_h/2 m_\pm)^2$ and 
$g_\pm = \lambda_1 v^2 / (2m_\pm^2)$ in the last term.  The effective
coupling strength is therefore $\kappa_\gamma = |A/6.5| \in [0.72,\, 1.14]$
using constraints from ATLAS and CMS \cite{Higgs_constraints}.  We then
find that the Higgs potential coupling $\lambda_1$ is bounded by
\be
	-0.7 < \lambda_1 < 1.4
\ee
for $m_\pm = 100\,$GeV.

\section{Charge transformation flavor ansatz}
\label{new}

Two Higgs doublet models have had a long history of proposed mechanisms
to control FCNCs, starting with that of 
Glashow and Weinberg \cite{Glashow:1976nt}, where up and down quarks
are restricted to couple to different Higgs doublets. More recent
ideas include the Cheng-Sher
texture \cite{Cheng:1987rs}, 
MFV \cite{Chivukula:1987py,D'Ambrosio:2002ex,Cirigliano:2005ck}
and alignment \cite{Pich:2009sp,Celis:2012dk}.  The ansatz we suggest
is distinct from these, and takes the form
\be
	\rho_u^\dagger V = \eta\,U\, V\rho_d
\label{CTM}
\ee
where $\eta\lesssim 1$ 
and $U$ is a diagonal unitary matrix, 
whose first element is 
$U_{11}= -1$, while the second is $U_{22}= +1$.  The third
element $U_{33}$ is not yet determined by experimental constraints.
We note that if $U_{33}\cong -1$ then $U$ would be special unitary.
The structure (\ref{CTM}) must of course be 
supplemented by a choice of entries for $\rho_d$ from 
which $\rho_u$ can be computed, or {\it vice versa}.  

Let us comment on the 
general utility of
this ansatz for the $B$ decays of interest.  The signs chosen for $U_{11}$ and $U_{22}$ are such that
the relations (\ref{hint1},\ref{hint2}) are satisfied.  The effect of the
sign difference between $U_{11}$ and $U_{22}$ can be understood by using
(\ref{CTM}) to eliminate $\rho_u^\dagger V$ from the 	
charged current interactions of the 
quarks with $H^+$, which then take the form
\be
\label{chargedLq}
	{\cal L} =
	 - H^+ \bar u_i\,(P_{\sss R}-\eta\, U_{ii} P_{\sss L}) (V\rho_d)_{ij} \,d_j
		+{\rm h.c.}
\ee
If $\eta\, U_{11}=-1$, the projection operators combine as $P_{\sss R}+P_{\sss L}=1$ for the
coupling to up quarks, forming a pure scalar $\bar u b$ that cannot
interpolate between the pseudoscalar $B^+$ meson and the vacuum, hence giving
no contribution to $B\to\tau\nu$ decay.  For $\eta\, U_{22}=+1$, the
combination $P_{\sss R}-P_{\sss L}=\gamma_5$ is pure pseudoscalar, in agreement
with the sign difference in the fit result 
$C^{cb}_{S_R}\cong -C^{cb}_{S_L}$ \cite{Freytsis:2015qca} for $B\to D\tau\nu$.  

The value of $\eta$ is independently determined by either of the
two anomalous measurements.  Using 
(\ref{BDfits2a}-\ref{BDfits2b}) and (\ref{Btaunu_target})
respectively, we find that
\be
	\eta \cong \left\{{0.78,\quad R(D^{(*)})\atop 0.83,\quad
	B\to\tau\nu}\right.
\label{etaeq}
\ee
It is encouraging that these two estimates are consistent within
the experimental errors.  We adopt the compromise $\eta=0.8$
in the following.  
We note that in the limit $\eta=1$, the excess in 
$B\to D^{(*)}\tau\nu$ is completely in the vector $D^*$ channel and absent from the
$D$ final states, because of the parity of the $\bar c\gamma_5 b$ 
pseudoscalar coupling.  By letting $\eta\neq 1$, this charged current
coupling acquires
a scalar component interpolating to pseudoscalar $D$ final states as well.
The current data are consistent with most of the anomaly being in the $D^*$ channel
since the error bars are smaller there (see table \ref{tab:anomalies}).

The relation (\ref{CTM}) at first sight looks peculiar, since  it
relates two flavor symmetry breaking effects, associated with quarks
of opposite charges.  For convenience we give it the name of ``charge
transformation" (CT) mechanism.   In section \ref{uvmodel} we will
show that such a structure can reasonably arise from a more
fundamental theory of flavor.   For now we will take it as a working
hypothesis and check whether it is sufficient to  help control 
FCNC's, in conjunction with some specific choices of $\rho_d$
couplings.

In section \ref{numerical} we will allow for all elements of $\rho_d$
to be nonzero, consistent with a wide variety of experimental
constraints. Here we make the approximation of  real-valued
$\rho_d$ (as well as CKM matrix) so that there are only nine
parameters in $\rho_d$.  The fact that there exists a solution that
can satisfy many more than nine constraints (not all of which are
upper bounds because of the anomalies) is striking. Moreover we
will show that there is no need for fine tuning of the parameters.

\section{FCNC constraints: meson mixing}
\label{constraints}

Although the anomalies in question can be accounted for with 
only the 
$\rho_d^{bb}$ element dominating in $\rho_d$ (and $\rho_d^{sb}\sim
0.02\, \rho_d^{bb}$), naturalness
demands that we consider nonvanishing values of the other entries.
Neutral meson mixing ($K^0$-$\bar K^0$, $D^0$-$\bar D^0$, 
$B^0$-$\bar B^0$, $B_s$-$\bar B_s$) provides strong constraints
on their sizes.  In this section we determine the tree-level and 
one-loop predictions of the model in the presence of general
couplings.

\subsection{Neutral meson mixing: generalities}

The new Higgs bosons induce contributions to neutral meson oscillations.
At the quark level, they can be described by an 
effective Hamiltonian in which the bosons have been integrated out.  In
general it can contain a number of operators with different Lorentz and color
structure.  Even though tree-level exchanges only produce two of these
operators, at one loop two additional ones are also generated.

The most general effective Hamiltonian for neutral meson mixing is
\be
	H = \sum_{ij}\left(\sum_{k=1,5}C^{ij}_k Q^{ij}_k + 
	\sum_{k=1,3} \tilde C^{ij}_k \tilde Q^{ij}_k\right),
\ee
where the flavor indices run over $ij=sd,cu,bs,bd$ (also
denoted $K$, $D$, $B_s$, $B_d$ respectively) and the 
operators are 
\bea
	Q_1^{ij} &=& (\bar q_{L,i}^\alpha \gamma^\mu q^\alpha_{L,j})\,(\bar
				q^\beta_{L,i}\gamma_\mu q^\beta_{L,j})\\
	Q_2^{ij} &=& (\bar q^\alpha_{R,i} q^\alpha_{L,j})\,(\bar q^\beta_{R,i} q^\beta_{L,j})\\
	Q_3^{ij} &=& (\bar q^\alpha_{R,i} q^\beta_{L,j})\,(\bar q^\beta_{R,i} q^\alpha_{L,j})\\
	Q_4^{ij} &=& (\bar q^\alpha_{R,i} q^\alpha_{L,j})\,(\bar q^\beta_{L,i} q^\beta_{R,j})\\
	Q_5^{ij} &=& (\bar q^\alpha_{R,i} q^\beta_{L,j})\,(\bar q^\beta_{L,i} q^\alpha_{R,j}).
\eea
Here $\alpha,\beta$ are colour indices, and $\tilde Q_k$ are related to $Q_k$ by taking
$R\leftrightarrow L$.  The coefficients of the latter are experimentally
constrained at the same level as the 
ones without tildes. 

Integrating out the neutral scalars, we obtain the coefficients
\bea
C_{2}^{ij} &=& \sum_{\phi= h,H,A}\frac{(y^\dagger_{\phi})_{ij}^2}{2 \, m_\phi^2},  \\
\tilde{C}_{2}^{ij} &=& \sum_{\phi= h,H,A}\frac{(y_{\phi})_{ij}^2}{2 \, m_\phi^2},  \\
C_{4}^{ij} &=& \sum_{\phi= h,H,A}\frac{(y^\dagger_{\phi})_{ij} \, (y_{\phi})_{ij}}{2 \, m_\phi^2}.
\eea
If $H$ and $A$ are nearly degenerate as we envision in our model, then
there is strong destructive interference between their contributions
to $C_{2},\tilde{C}_{2}$.  This can be understood in terms of the original
complex fields $H\pm iA$ from the fact that the $\langle(H\pm iA)^2\rangle$
propagator vanishes when $H$ and $A$ are degenerate.  On the other hand
there is no such cancellation for $C_{4}^{ij}$, so it provides the most
stringent constraints, unless one of $(y_{\phi})_{ij}$ or
$(y^\dagger_{\phi})_{ij}$ vanishes. However naturalness favors roughly
symmetric Yukawa matrices, as we will show, so that $C_{4}^{ij}$ is not suppressed in this
way.

\subsection{Tree-level constraints on mixing}

New tree-level contributions to neutral meson
mixing are mediated by the neutral Higgs bosons.
The $C_2^{ij}$ coefficients get contributions of opposite
signs from the CP-even and odd boson exchanges.   
 Using the mass relation
(\ref{massrels}), they can be reorganized into the form 
\bea
\label{C2coeffs}
\tilde{C}_{2}^{ij} &=&
\frac{(\rho_d^{ij})^2}{2} \left[{\cba^2\over m_h^2} - 
	{\cba^2\over m_A^2}  
	+ {\cba^2 m_h^2 - 2\lambda_3v^2\over m_A^2\, m_H^2} 
 \right] \\
{C}_{2}^{ij} &=&
\frac{(\rho_d^{ji*})^2}{2} \left[{\cba^2\over m_h^2} - 
	{\cba^2\over m_A^2}  
	+ {\cba^2 m_h^2 - 2\lambda_3v^2\over m_A^2\, m_H^2} 
 \right]\nn
\eea
The terms proportional to $\cba^2$ are negligible
if $\lambda_3$ is not too small.
Later we will find that $\lambda_3\cong 10^{-3}$ can be consistent
with technical naturalness, so we adopt this value in what follows.
For $m_H \cong  m_A \cong 100\,$GeV,
the constraints on the coefficients become \cite{Bona:2007vi,Gedalia:2009kh}
\bea
\label{tree-mixing2}	
	\bar\rho_d^{sd}&<& \left\{{1.3\cdot 10^{-4}
	\atop 9.3\cdot 10^{-6}}\right\},\quad
	\bar\rho_d^{cu} < \left\{{5.1\cdot 10^{-4}
	\atop 1.2\cdot 10^{-4}}\right\}\nonumber\\
	\bar\rho_d^{bd}&<& 1.1\cdot 10^{-3},\qquad\quad
	\bar\rho_d^{bs} < 9.6\cdot 10^{-3}
\eea
where $\bar\rho_d^{ij}$ stands for either 
$|\rho_d^{ji}|$ or $|\rho_d^{ij}|$.  For $K$ $(sd)$ and 
$D$ ($cu$), we show the separate limits from the real (upper) and 
imaginary (lower)
parts of $C_2^{ij}$. In our fits we will impose the more stringent
ones.   
These limits scale
as $(m_H/100{\rm\,GeV})(\lambda_3/10^{-3})^{-1/2}$.  

In the limit of small Higgs mixing and nearly degenerate $H$ and $A$, the
$C_4^{ij}$ coefficients take the form
\be
	C_4^{ij} \cong {\rho_{q}^{ij} \rho_q^{ji*}\over m_H^2}
\label{C4eq}
\ee
where $q=d$ for $K$, $B_d$, $B_s$ and $q=u$ for $D$ mixing.  Unlike $C_2$,
they are not suppressed by $\cba$ or $\lambda_3$.  Again for $m_H = 100\,$GeV, we have
the upper limits
\bea
\label{tree-mixing4}
	\sqrt{|\rho_d^{ds}\rho_d^{sd}|} &<& \left\{{6.0\cdot 10^{-6}
	\atop 4.2\cdot 10^{-7}}\right\},\ 
	\sqrt{|\rho_u^{uc}\rho_u^{cu}|} < \left\{{2.4\cdot 10^{-5}
	\atop 1.0\cdot 10^{-5}}\right\}\nonumber\\
	\sqrt{|\rho_d^{db}\rho_d^{bd}|} &<& 4.6\times 10^{-5},\quad
	\sqrt{|\rho_d^{sb}\rho_d^{bs}|} < 4.0\times 10^{-4}\quad\quad
\eea
They scale as $m_H/(100\,\rm{GeV})$.  By comparison of 
(\ref{tree-mixing2}) and (\ref{tree-mixing4}) we see that
$C_4$ gives more stringent constraints than $C_2$ if the coupling
matrices $\rho_q^{ij}$ are symmetric, which will be approximately
true in our later determination.

\subsection{One-loop contributions to mixing}

At one loop, the charged and neutral Higgs bosons give contributions to
neutral meson mixing  that are higher order in $\rho_{u,d}$, except for
loops involving $W^\pm$ exchange.   
We start with box diagrams involving the exchange of two
scalars, followed by exchange of one $H^\pm$ and a $W$ boson.
For mesons containing down-type quarks we find
\bea
	C_1^{ij} &=& {1\over 128\pi^2}\left( {(\rho_d^\dagger\,\rho_d)_{ij}^2
	\over m_\pm^2} + {(\rho_d\,\rho_d^\dagger)_{ij}^2
	\over m_H^2}\right)\nonumber\\
	\tilde C_1^{ij} &=& {(\rho_d^\dagger\,\rho_d)_{ij}^2\over 128\pi^2}
	\left( {1\over m_\pm^2} + {1\over m_H^2}\right)\\
	C_2^{ij} = \tilde C_2^{ij} = \sfrac14 C_4^{ij} &=& 
	[(V\rho_d)^\dagger_{it}(V\rho_d)_{tj}]^2 {m_t^2(\ln {m_\pm^2\over m_t^2}-2)
	\over 64\pi^2\, m_\pm^4}\nonumber\\
	C_5^{ij} &=& {(\rho_d^\dagger\rho_d)_{ij}\over 32\pi^2}\left({(\rho_d\rho_d^\dagger)_{ij}
	\over m_H^2} + {(\rho_d^\dagger\rho_d)_{ij}
	\over m_\pm^2}\right)\nonumber
\eea
while for $D^0$ mesons
\bea
	C_1^{uc} &=& {1\over 128\pi^2}\left( {(V\rho_d\,\rho_d^\dagger V^\dagger)_{uc}^2
	\over m_\pm^2} + {(V \rho_d^\dagger\,\rho_d V^\dagger)_{uc}^2
	\over m_H^2}\right)\nonumber\\
	\tilde C_1^{uc} &=& {(V\rho_d\,\rho_d^\dagger V^\dagger)_{uc}^2\over 128\pi^2}
	\left( {1\over m_\pm^2} + {1\over m_H^2}\right)\\
	C_5^{uc} &=& -{(V\rho_d\,\rho_d^\dagger V^\dagger)_{uc}^2\over 32\pi^2\,m_\pm^2}
	- {(V\rho_d\,\rho_d^\dagger V^\dagger)_{uc}(V\rho_d^\dagger\,\rho_d V^\dagger)_{uc}
	\over 32\pi^2\,m_H^2}\nonumber
\eea
We omit the $C_{2,4}^{uc}$ coefficients that are suppressed by $m_b^2$.

The box diagrams containing $H^\pm W^\mp$ exchange give rise to 
\bea
	C_1^{ij} &=& {g_2^2 V_{tj} V^*_{ti} (V\rho_d)_{tj} (V\rho_d)^*_{ti}\over 128\pi^2\,m_\pm^4}
	\,m_t^2\left(\ln {m_\pm^2\over m_t^2}-2\right)\nonumber\\
	C_4^{ij} &=& {g_2^2 (\rho_d^\dagger)_{ij}(\rho_d)_{ij}\over 64\pi^2\,m_\pm^2}\nonumber\\
\eea
for down-quark type mesons, while for $D^0$ mesons
\be
	C_4^{uc} = {g_2^2(V\rho_d^\dagger V^\dagger)_{uc} (V\rho_d V^\dagger)_{uc}\over
	64\pi^2\, m_\pm^2}
\ee
and we neglect the $m_b^2$-suppressed contribution to $C_1^{uc}$.  

These loop contributions turn out to be much smaller than the tree-level 
ones previously considered; in the numerical fit of section
\ref{numerical} they are at most a factor of $\sim 10^{-3}$ below the
upper limits.  
But since they depend upon different combinations of the $\rho$ 
couplings, which are 
hierarchical, it was not {\it a priori} obvious  that they should be 
negligible.

\section{Rare decays and $(g-2)_\tau$}
\label{decays}

Our model predicts a variety of rare decays
beyond those already considered in section \ref{anomalies}, and a new
contribution to the anomalous magnetic moment of the $\tau$.  Although they are 
potentially constraining, most of them  
turn out to be less so than tree-level meson
mixing.  The loop enhancement of $Z\to\ttpm$ (and
$Z\to\nu_\tau\bar\nu_\tau$) is the most
important of these since it sets the limit on how large the
$\rhoett$ coupling can be, which is central to the explanation of
the $\tau$ anomalies.  The prediction for $\tau\to\pi^-\nu$ is close to
the experimental limit for this decay, while 
the NP amplitude of $b\to s\gamma$ is 
the next most significant, a factor of four below the experimental limit.

\begin{figure}[b]
\centerline{
\includegraphics[width=0.9\columnwidth]{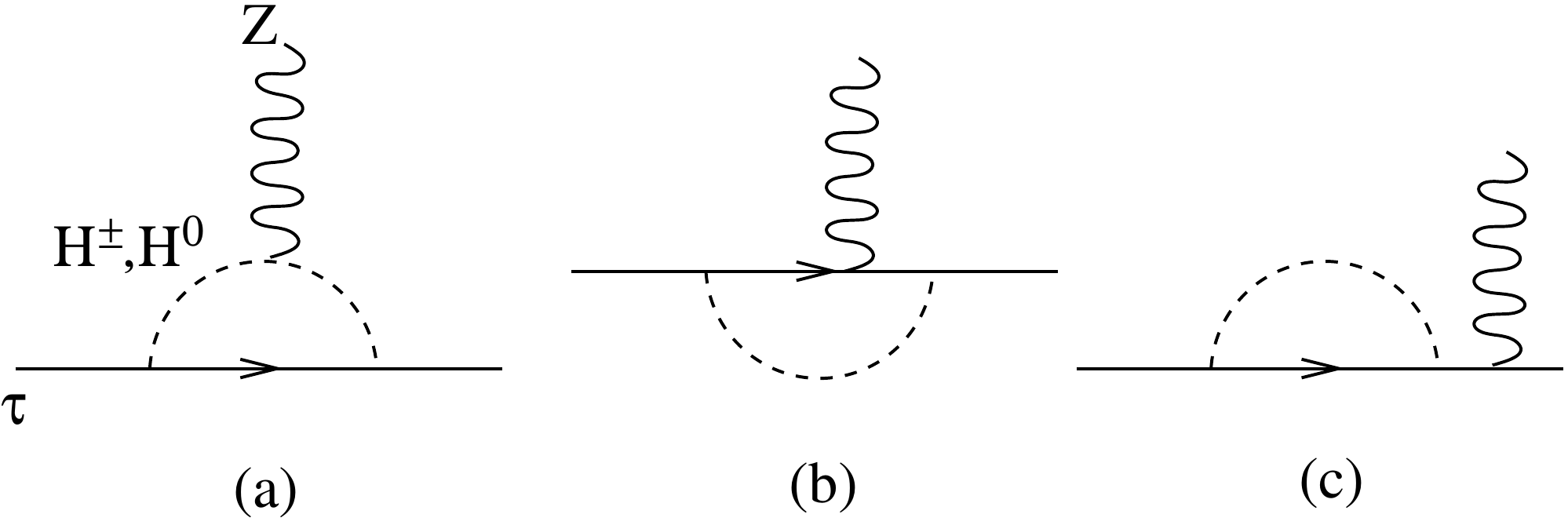}}
\caption{Diagrams contributing to $Z\to\ttpm$.}
\label{Zdecay}
\end{figure}
\begin{figure}[t]
\centerline{
\includegraphics[width=\columnwidth]{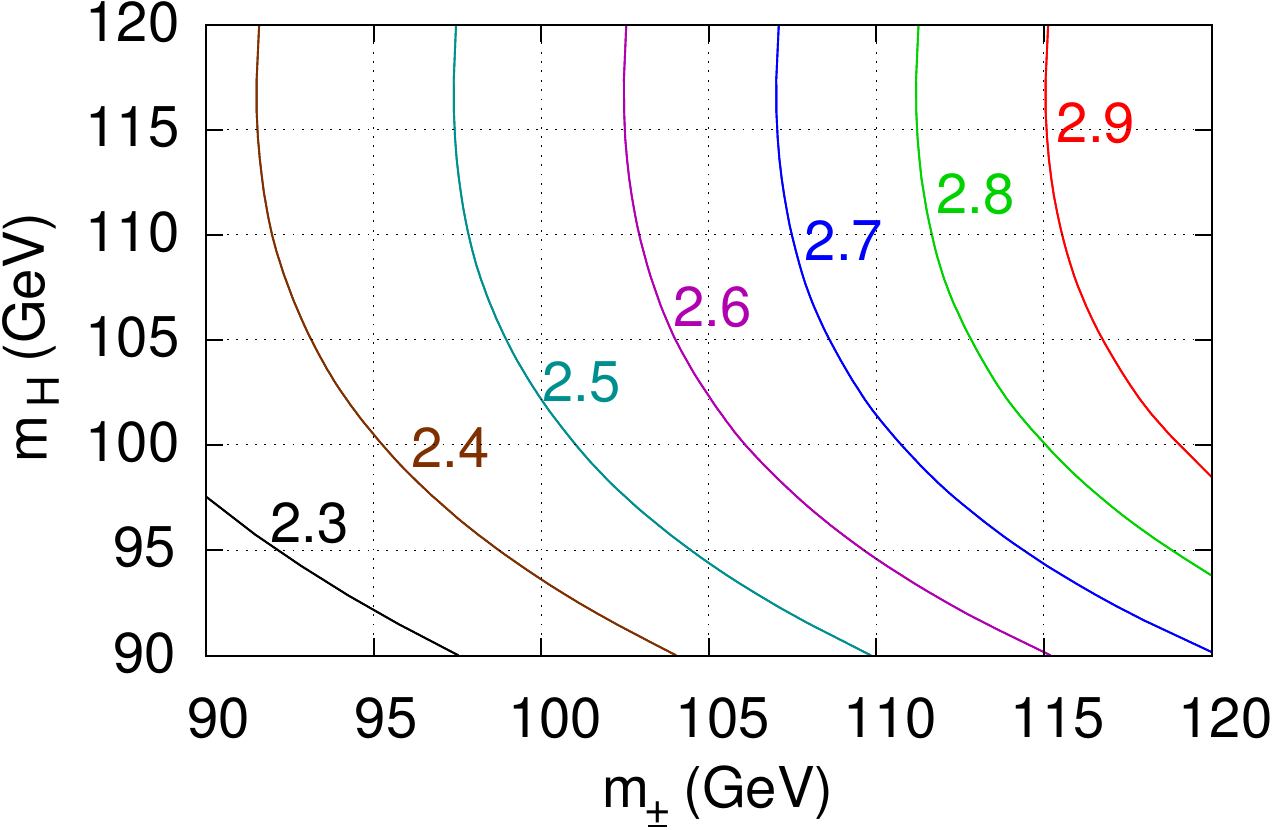}}
\caption{Contours of upper bound on $|\rhoett|$ 
from lepton flavor universality of $Z\to\ell^+\ell^-$ decays,
in the plane of neutral versus charged Higgs boson masses.}
\label{Ztautau}
\end{figure}

\subsection{$Z\to \ttpm$}
\label{Zll}

The coupling $\rhoett$ introduces lepton universality violation
in $\Gamma(Z\to \bar{\ell} \, \ell)$, when comparing
 $\ell = \tau$ to
$\ell = e,\mu$. Such deviations are constrained by LEP, which has reported  \cite{PDG}
\bea\label{leptonuni}
R_{\tau/e} = \frac{\Gamma(Z \to \tau^+ \tau^-)}{\Gamma(Z\to e^+ e^-)}
 = 1.0019\pm 0.0032
\eea
The new contributions from exchange of the heavy charged and neutral scalars are 
shown in fig.\ \ref{Zdecay}.  These one-loop diagrams
give the effective interaction term for the right-handed component of
$\tau$ coupling to $Z$:
\bea
\label{leff_tauR}
\mathcal{L}_{\rm eff,\tau_R} &=& -(g_{\tau_R} + \delta g_{\tau_R}) 
\,(\bar\tau_R\gamma_\mu\tau_R)\, Z^\mu \\
	 g_{\tau_R} &=&  g_Z\,\sw^2,\quad
	g_Z = {e\over \cw\sw}\nonumber\\
	\delta g_{\tau_R} &=& -{|\rhoett|^2 
	g_Z\over 32\pi^2}\,F_{\tau_R},
	\nonumber\\
	F_{\tau_R} &=& \sfrac12\left(F_a^0 -(1-2\sw^2)F_b^0 + 
	\sw^2 F_c^0\right)\nonumber\\
	&+& \sfrac12\left(-(1-2\sw^2)F_a^\pm + F_b^\pm + \sw^2 F_c^\pm
	\right)\nn
\eea
where $\cw = \cos\theta_W$ and $\sw = \sin\theta_W$.  In the limit of vanishing lepton
masses,  the loop integrals involving the neutral Higgs $H^0$ are given by
\bea
\label{loopfns}
	F_a^0 &=& 2\int_0^1 dx\int_0^{1-x}dy\left(\ln {(x+y)m_H^2 - xy\, m_Z^2\over\mu^2}
	\right)\nonumber\\
	F_b^0 &=& 2\int_0^1 dx\int_0^{1-x}\!\!\!\!\!\!dy\,\left(
	\ln{M_b^2\over \mu^2}+{x y\,m_Z^2\over M_b^2} + 1\right)
	\nonumber\\
	& &\left[\hbox{with}\ M_b^2 = (1-x-y)\,m_H^2 -xy\, m_Z^2\right]\nonumber\\
	F_c^0 &=& 1 - 2\ln {m_H^2\over \mu^2}
\eea
while those involving $H^\pm$ (denoted by $F^\pm_i$) are the same 
but with $m_H^2\to
m_\pm^2$.  
We have neglected terms of order $\cba^2$ in the Higgs mixing.
Dependence on the renormalization scale $\mu$ drops out in $F_{\tau_R}$
and $F_{\tau_L}$ (below).

The analogous expressions to (\ref{leff_tauR}) for the couplings of
$\tau_L$ are given by
\bea
\label{leff_tauL}
\mathcal{L}_{\rm eff,\tau_L} &=& -(g_{\tau_L} + \delta g_{\tau_L}) 
\,(\bar\tau_L\gamma_\mu \tau_L)\, Z^\mu \\
	g_{\tau_L} &=& -\sfrac12 g_Z(1- 2\sw^2)\nonumber\\
	\delta g_{\tau_L} &=& -{|\rhoett|^2 g_Z\over 32\pi^2}
	F_{\tau_L},\nonumber\\
	F_{\tau_L} &=& -\sfrac12 F_a^0 + \sw^2\, F_b^0 -
\sfrac14(1-2\sw^2)\, F_c^0\nn
\eea

Writing $R_{\tau/e} = 1 + \Delta R_{\tau/e}$, 
the predicted value of the deviation is
\bea
\label{deltaR}
\Delta R_{\tau/e} &=& 2\,{g_{\tau_R}\delta g_{\tau_R} + 
	g_{\tau_L}\delta g_{\tau_L}\over g_{\tau_R}^2 + g_{\tau_L}^2}
	\\
 	&=& {|\rhoett|^2\over 8\pi^2}\left(
	{-2\sw^2 F_{\tau_R} + (1-2\sw^2) F_{\tau_L}\over
	4\sw^4 + (1 - 2\sw^2)^2}\right) \nn
\eea
We take the 1$\sigma$ experimental upper limit which gives
\be
\label{dRlimit}
	-0.0013 < \Delta R < 0.0051
\ee to obtain the upper bounds on 
$|\rhoett|$ shown in fig.\ \ref{Ztautau}.
(A similar calculation for more general boson masses
was carried out in ref.\ \cite{Abe:2015oca}.)

\subsection{$Z\to\nu_\tau\bar\nu_\tau$}
Similar diagrams to those in fig.\ \ref{Zdecay} contribute to the
amplitude for $Z\to\nu_\tau\bar\nu_\tau$ decay.  We find that 
the perturbation to the tree-level coupling in analogy to
(\ref{leff_tauR},\ref{leff_tauL}) is given by
\bea
\label{leff_nutau}
\mathcal{L}_{\rm eff, \nu_\tau} &=& -(g_{\nu_\tau} + \delta g_{\nu_\tau}) 
\,(\bar\nu_\tau\gamma_\mu \nu_\tau)\, Z^\mu\\
	g_{\nu_\tau} &=& \sfrac12 g_Z\nonumber\\
	\delta g_{\nu_\tau} &=& -{|\rhoett|^2 g_Z\over 32\pi^2}
	F_{\nu_\tau},\nonumber\\
	F_{\nu_\tau} &=& \sfrac12(1-2\sw^2)F_a^\pm + \sw^2\, F_b^\pm +
	\sfrac14\, F_c^\pm\nn
\eea

The branching ratio to invisible decays is changed by 
\be
	{\Delta\Gamma_{\rm inv}\over \Gamma_{\rm inv}} = 
	{2g_{\nu_\tau}\delta g_{\nu_\tau}\over 3\, g_{\nu_\tau}^2}
	= -{|\rhoett|^2 F_{\nu_\tau}\over 24\pi^2}
\ee
For the fiducial values $\rhoett=2.5$, $m_\pm=100{\rm\,GeV}$ that we will 
adopt, this leads to an increase $\Delta\Gamma_{\rm inv}=1.1{\rm\,MeV}$
in the invisible width of the $Z$. This is close to but consistent with 
the combined LEP upper limit $\Delta\Gamma_{\rm inv} < 2{\rm\,
MeV}$ \cite{ALEPH:2005ab} at 95\% c.l..

\subsection{$W\to\tau\nu$}
Distinct from the tree-level $H^\pm$ decays  that might
have faked $W\to\tau\nu$ events at LEP, discussed in section 
\ref{Wtaunu},
there is an actual perturbation to 
the amplitude for $W\to\tau\nu$ from loops analogous to those in 
fig.\ \ref{Ztautau}. In this case, diagrams of type
(b) do not contribute because they require a chirality flip leading to
suppression by $m_\tau$, since $W$ couples only to left-handed particles.  The remaining diagrams give a fractional
correction to the $W\tau\nu$ coupling of
\be
	{\delta g_{W\tau\nu}\over g_{W\tau\nu}}
	= -{|\rhoett|^2\over 64\pi^2}
	\left[2 F_a^\pm + \sfrac12(F_c^0+F_c^\pm)\right]
\ee
with the loop functions in brackets evaluating to $\sim -0.1$ for the Higgs
boson masses of interest.  For $\rhoett=2.5$, this leads to a fractional increase in the 
branching ratio of $0.2\%$, which is the same as the experimental error
\cite{PDG}.  This contribution, while not enough by itself, goes in the right
direction  and could work in combination
with the $H^\pm$ decays to explain the the observed excess.

\subsection{$\tau$ anomalous magnetic moment}

The anomalous magnetic moment of the $\tau$ is at present
only weakly constrained, $-0.052 < \Delta a_\tau < 0.013$
\cite{PDG}.  At one loop, the leading contribution in our model
comes from neutral $H$ exchange 
(see for example ref.\ \cite{Ilisie:2015tra}),
\be
	\Delta a_\tau = {|\rhoett|^2\, m_\tau^2\over 8\pi^2
	\, m_H^2}\, F\left(m_\tau\over m_H\right) \cong 2\times 10^{-5} 
	\left(100{\rm\, GeV}\over m_H\right)^2
\ee
where the loop function evaluates to $F\cong 7$.  The analogous 
contribution from charged Higgs exchange has a much smaller loop
function $\cong -0.2$.  

Frequently the dominant contribution to such processes in 2HDMs is the two-loop
Barr-Zee (or Bjorken-Weinberg) \cite{Bjorken:1977vt,Barr:1985ig}
diagram with a top quark or other particle in one of the loops. 
We find that indeed the contribution from the top quark loop
exceeds the one-loop contribution, giving
\be
	\Delta a_\tau = {\alpha\rhoett\rho_u^{tt}\over
	8\pi^3}\,{\cal F}^{(1)}[(m_t/m_H)^2] \lesssim 7\times 10^{-4}
\label{barr-zee}
\ee
where ${\cal F}^{(1)}[(m_t/m_H)^2] = -1.14$ for $m_H=100\,$GeV,
using the notation of
ref.\ \cite{Ilisie:2015tra}, and we took $\rho_u^{tt}\lesssim 0.08$ from
eqs.\ (\ref{rhodbb_const},\ref{etaeq}).  Although this is much smaller
than the current experimental bound, it is two orders of magnitude
larger than the SM prediction \cite{Burger:2015oya}.  We find that the
other Barr-Zee diagrams are smaller, contributing $10^{-6}$ from the
$\tau$ loop analogous to (\ref{barr-zee}) and $10^{-5}$ from the 
diagrams with $t,b,\nu,H^\pm,W^\pm$ in the loops 
($\Delta a_\tau^{(4)}$ in the notation
of \cite{Ilisie:2015tra}).

\subsection{Hadronic decays $\tau\to \pi^-\nu,\, K^-\nu$ and
$D_s\to\tau\nu$} 
\label{taudecays}
Charged Higgs exchange contributes to hadronic decays of the $\tau$,
the simplest of which are $\tau\to\pi^-\nu$ with 
branching ratio $B=(10.83\pm 0.06)\%$
and $\tau\to K^-\nu$ with $B=(7.00\pm 0.10)\times 10^{-3}$ \cite{PDG}.
The amplitude for $\tau\to K^-\nu$ was already given in eq.\
(\ref{MtauKnu}).  Using the CT flavor ansatz (\ref{CTM}) we have 
$(V\rho_d +\rho_u^\dagger V)_{us} = (1-\eta)(V\rho_d)_{us}$.
For $\tau\to\pi^-\nu$, one replaces
$(V\rho_d)_{us}\to (V\rho_d)_{ud}$, $m_s\to m_d$, 
$m_K\to m_\pi$ and $f_K\to f_\pi$.

In both decays, the NP contribution
interferes with that of the SM. If we assume that the hint for new
physics in $\tau\to K\nu$ discussed in section \ref{tauKnu} is just due to
a statistical fluctuation, then by demanding that the extra
contribution to the branching ratio does not exceed the experimental error, 
we find the constraints
\be
	|(V\rho_d)_{us}| < 7\times 10^{-4},\quad
	|(V\rho_d)_{ud}| < 1\times 10^{-3}
\label{taudecaybounds}
\ee
assuming that $\rhoett=2.5$ and $m_\pm = 100\,$GeV.
We note that this constrains couplings $\rho_d^{qd}$ and 
$\rho_d^{qs}$ different from those ($\rho_d^{sb}$ and $\rho_d^{bb}$)
required to explain the $B$ decay anomalies.
In the fit to be described below (section \ref{numerical}), we obtain
$(V\rho_d)_{us} = 2\times 10^{-4}$, $(V\rho_d)_{ud} = 9.5\times 10^{-4}$.
The NP contribution to $\tau\to\pi^-\nu$ is therefore close to the limit.

The matrix element for $D_s\to\tau\nu$ is 
\be
	{\cal M}_{D_s\to\tau\nu} =
(1+\eta)\,{\rhoett(V\rho_d)_{cs}\, f_{D_s}\, m_{D_s}^2
	\over 2\, m_\pm^2\, (m_c+m_s)}\, 
	(\bar u_\tau P_{\sss L} u_\nu)
\ee
where $f_{D_s} = 0.248\,$GeV \cite{Davies:2010ip}.  Using the observed
branching ratio $(5.55\pm 0.24)\%$ \cite{PDG} we obtain the bound
\be
	|(V\rho_d)_{cd}| < 1\times 10^{-3}
\ee
The value from our fit, $(V\rho_d)_{cd}=-2\times 10^{-4}$, is consistent.

\subsection{$b\to s\gamma$}
The off-diagonal couplings in $\rho_d$ and $V\rho_d$ introduce new contributions
to $b\to s\gamma$ at one loop, which is encoded by the effective 
Hamiltonian \cite{Lunghi:2006hc}
\bea
\label{bsgH}
	H &=& -{4 G_F\over \sqrt{2}}\, V_{tb} V^*_{ts}\, 
	 \left( C_7\, {\cal O}_7 + C_7'\, {\cal O}_7'\right)\\
	{\cal O}_7\, ({\cal O}_7') &=& { e\, m_b\over 16\pi^2}\,
	\bar s\, (\sigma_{\mu \nu} 
	P_{\sss R (L)})\, b
	\,F^{\mu\nu}\nonumber 
\eea
Because of operator mixing, one should also consider the analogous operators
${\cal O}_8$, ${\cal O}_8'$ for the chromomagnetic moments.
These have been computed for type I and II 
2HDMs \cite{Grinstein:1988me,Grinstein:1990tj} 
but not (as far as we can tell) for a general type III model.  
A full computation for
this case might be interesting for future study.  However most of the 
contributions appearing in our model can be inferred from the earlier
calculations by transcribing the right- and left-handed couplings of the 
charged Higgs from the type I/II models, 
\be
\label{bsgL}
	{\cal L} = (4G_F/\sqrt{2})^{1/2} \bar u\left(
	\xi m_u V P_L - \xi' V m_d P_R \right) d
\ee
where $m_{u,d}$ are the quark mass matrices.  The dominant contributions to
the one-loop charged Higgs diagrams in our model can be estimated by 
taking
\bea
\label{xixip}
	\xi^2 &\to& {1\over 2m_t^2 G_F}\, {(\rho_u^\dagger V)_{ts}\over V_{ts}}
	\,  {(\rho_u^\dagger V)_{tb}\over V_{tb}} \nn\\
		&=&  45\,U_{tt}(V\rho_d)_{ts}(V\rho_d)_{tb}\\
	\xi\xi' &\to& {1\over 2m_t\, m_b G_F}\, {(\rho_u^\dagger V)_{ts}
	\over V_{ts}}
	\,  {(V\rho_d)_{tb}\over V_{tb}} \nn\\
	&=& 2.3\times 10^3\, (V\rho_d)_{ts}(V\rho_d)_{tb}
\eea
in the Wilson coefficients \cite{Grinstein:1988me,Aliev:1997uz}
\bea
\label{C7C8}
	C_7 &=& \xi\xi'\left({-3y^2+2y\over 6(y-1)^3}\ln y + 
	{3y-5y^2\over 12(y-1)^2}\right)\\
	&+& \xi^2 \left({-3y^3+2y^2\over 12(y-1)^4}\ln y + 
	{-8y^3-5y^2+7y\over 72(y-1)^3}\right)\nn\\
	C_8 &=& \xi\xi'\left({y\over 2(y-1)^3}\ln y + 
	{y^2-3y\over 4(y-1)^2}\right)\nn\\
	&+& \xi^2 \left({y^2\over 4(y-1)^4}\ln y + 
	{-y^3+5y^2+2y\over 24(y-1)^3}\right)\nn
\eea
where $y = (m_t/m_\pm)^2$.  In (\ref{xixip}) we have indicated the
expressions following from our flavor ansatz (\ref{CTM}) that involve the
undetermined sign $U_{tt} = \pm 1$.  
In the type I/II models, $C_7'$ is smaller
than $C_7$ by a factor of $m_s/m_b$, but we do not expect that in our model
since there is no suppression of the right-handed couplings by $m_d$.
Instead, the primed coefficients are given by (\ref{C7C8}) after
interchanging $\rho_u^\dagger V \leftrightarrow V\rho_d$ in (\ref{xixip}). 
With our flavor ansatz (\ref{CTM}), this implies that $\xi^2$ becomes larger
by the factor $1/\eta^2 = 1.56$ while $\xi\xi'$ remains the same.

Recent constraints on $C_7$ and $C_7'$ (by which we always mean the NP
contributions) at the scale of $m_b$ have been
determined by ref.\ \cite{Descotes-Genon:2015uva},
\bea
	C_7(m_b) &\in& [-0.055,\,0.02],\nn\\
  C'_7(m_b) &\in& [-0.03,\,0.065]
\eea
at $2\,\sigma$.  The 
coefficients (\ref{C7C8}) evaluated at the weak scale must be run down 
to $m_b$ \cite{Grinstein:1990tj},
\bea
	C_7(m_b) &=& \eta^{16/23}C_7 + \sfrac83(\eta^{14/23} -
\eta^{16/23})C_8\nn\\
	&\cong& 0.6\, C_7 + 0.1\, C_8
\eea
at leading order in QCD corrections, 
where $\eta = \alpha_s(m_W)/\alpha_s(m_b)\cong 0.5$.  The primed coefficients
run in the analogous way.
The numerical fit of section \ref{numerical} yields $C_7(m_b)\cong C_7'(m_b)
\cong 4.9\times 10^{-3}$, four times below the limit for $C_7$.

There are also Barr-Zee two-loop contributions that we find to be much
smaller.  For example the diagram with a top quark loop and neutral $H^0$
exchange generates \cite{Davidson:2010xv,Sierra:2014nqa}
\be
	C_7 = {\sqrt{2}\, N_c\, Q_b\, Q_t^2\,\alpha\, \rho_d^{sb}\,\rho_u^{tt}\over
		8\pi\, m_b\, m_t\, G_F\, V_{tb}\, V_{ts}} f(m_t^2/m_H^2) \lesssim
10^{-4}
\ee
where the loop function $f(m_t^2/m_H^2)\cong -1$.

\subsection{$s\to d\gamma$}

For the radiative decays of lighter quarks, it is
not necessarily a good approximation to assume that 
the top quark contribution in the loop dominates, because
the relevant coupling $(V\rho_d)_{td}$ is CKM-suppressed, and
for $c\to u\gamma$ the dominant graph is from an internal
$b$ quark. For these decays we content ourselves with an
estimate based upon the analogous treatment of leptonic processes
$\tau\to\mu\gamma$ studied in 2HDMs \cite{Davidson:2010xv}, which
obtains the separate contributions from neutral as well as charged Higgs
exchange.  Defining the operator coefficients in the effective Hamiltonian
as 
\bea
\label{sdgH}
	{\cal H}_{\rm eff} &=& {Q_s\, e\, m_s\over 2}\,\bar s\, \sigma_{\mu \nu} 
	(A^{sd}_{\sss R} P_{\sss R}+ A^{sd}_{\sss L} P_{\sss L}) b
	\,F^{\mu\nu}
\eea
we find
\bea
	A^{sd}_R = (A^{sd}_L)^*\!\!\! &=&\!\!\!\!\!\! \sum_{q=u,c,t}
	\!\! {Q_q\, m_q\over Q_s\, m_s}
	{(V\rho_d)^*_{qd}(V\rho_d)_{qs}\over 16\pi^2\, m_\pm^2}
	\,f\left(m_\pm\over m_q\right)\nn\\
\label{sdgamma}
	A^{sd}_R\ [A^{sd}_L]\! &=& \!\!\!\!\!\!\sum_{q=d,s,b}\!{m_q\over m_s}{\rho_d^{dq}\rho_d^{qs}\
	[(\rho_d^{qd}\rho_d^{qs})^*]\over 16\pi^2\, m_H^2} 
	\,f\left(m_H\over m_q\right)\nn\\
	A^{sd}_R\ [A^{sd}_L] &=& {\rho_d^{ds}\,y_s\,\cba\
	[\rho_d^{sd*}\,y_s\,\cba]\over 16\pi^2\, m_h^2} 
	\,f\left(m_h\over m_s\right)
\eea
where the loop function is $f(x)\cong \ln x^2-3/2$.
Our numerical fit values of the couplings implies
$|A^{sd}_{L,R}|\cong -2\times 10^{-5}\,$TeV$^{-2}$. 

The dipole operator gives rise to a hadronic matrix element 
\be
	\langle \pi^0|\bar s\sigma^{\mu\nu}d|K^0\rangle = 
	\left(p_\pi^\mu p_K^\nu - p_K^\mu p_\pi^\nu\right){\sqrt{2}\,f_T^{K\pi}\over 
	m_K + m_\pi}
\ee
with $f_T^{K\pi}=0.4$  \cite{Baum:2011rm}.  It vanishes for on-shell 
photons in the decay $K\to\pi\gamma$, but gives a nonvanishing
contribution to leptonic modes mediated by the off-shell photon.  
Because $A_L^{sd} = 
A_R^{sd*}$, it does not contribute to the CP-violating decay
$K_L\to \pi\ell^+\ell^-$, but it does contribute to 
$K_S\to \pi\ell^+\ell^-$ whose measured branching ratio is $(3\pm 1.5)\times 10^{-9}$
\cite{PDG}.  

Adapting results of ref.\ \cite{Becirevic:2000zi} for
$K_L$ decay, we find 
\be
	{\Delta B(K_S\to \pi^0 e^+e^-)\over 
	B(K^+\to \pi^0 e^+\nu_e)}
	 = \left(2\zeta e^2 Q_s m_s 
	\tilde B_T {\rm Re}(A_R^{sd})\over V_{us}G_F m_K\right)^2 
	{\tau(K_S)\over\tau(K^+)}
\label{kpiee}
\ee
where $\tilde B_T = 1.2$ and $\zeta$ accounts for the renormalization
of $A_R^{sd}$ between the scale $m_H$ and $\mu = 2\,$GeV where the
lattice matrix elements are computed.  Assuming that the
chromomagnetic moment $g_s \bar d\sigma_{\mu\nu} G^{\mu\nu} s$
gets generated with the same coefficient as the electromagnetic one
$Q_s e \bar d\sigma_{\mu\nu} F^{\mu\nu}s$ at the scale $m_H$ and
accounting for the mixing of these operators under renormalization,
$\zeta = \eta^2(1-3\cdot8(1-\eta^{-1})) = 2.7$, where $\eta =
(\alpha_s(m_H)/\alpha_s(m_b))^{2/23}
(\alpha_s(m_b)/\alpha_s(\mu))^{2/25} = 0.9$.  Eq.\ (\ref{kpiee})
then gives the new physics contribution 
$\Delta B(K_S\to \pi^0 e^+e^-)= 3\times 10^{-13}$, far below the
measured value.

\subsection{$c\to u\gamma$}

Proceeding similarly to the case of $s\to d\gamma$, 
the dipole operators for $c\to u\gamma$ get contributions to their coefficients given by
\bea
	A^{cu}_R = (A^{cu}_L)^* \!\!\!&=&\!\!\!\!\!\! \sum_{q=d,s,b}
	\!{Q_q\, m_q\over Q_c\, m_c}
	{(V\rho_d)_{uq}(V\rho_d)^*_{cq}\over 16\pi^2\, m_\pm^2}
	\,f\left(m_\pm\over m_q\right)\nn\\
	A^{bs}_R\ [A^{cu}_L] \!\! &=& \!\!\!\!\!\! \sum_{q=u,c,t}
	\!{m_q\over m_c}{\rho_u^{uq}\rho_u^{qc}\
	[(\rho_u^{cq}\rho_u^{qu})^*]\over 16\pi^2\, m_H^2} 
	\,f\left(m_H\over m_q\right)\nn\\
	A^{cu}_R\ [A^{cu}_L] &=& {\rho_u^{uc}\,y_c\,\cba\
	[\rho_u^{cu*}\,y_c\,\cba]\over 16\pi^2\, m_h^2} 
	\,f\left(m_h\over m_c\right)
\label{cugamma}
\eea
The second of these (mediated by $H^0$ in the loop) is the largest, 
contributing $A_{R}^{cu}\cong 2\times 10^{-4}\,$
TeV$^{-2}$.  It is difficult to put precise constraints on this quantity
because of highly uncertain long-distance contributions to the observable
amplitudes.  Here we content ourselves with a comparison to the SM
short-distance contribution, estimated to be $A^{cu}_{SM} = 0.02\,G_F V_{us}
V_{cs}/(2\sqrt{2}\pi^2 Q_c) \cong 2\times 10^{-3}\,$TeV$^{-2}$ 
\cite{Greub:1996wn,Fajfer:2002bu}.  On this basis the new 
contribution appears to be sufficiently small, especially since the
observed $\Delta c=1$ decays are dominated
by the long-distance contributions.

\begin{figure}[t]
\centerline{
\includegraphics[width=0.9\columnwidth]{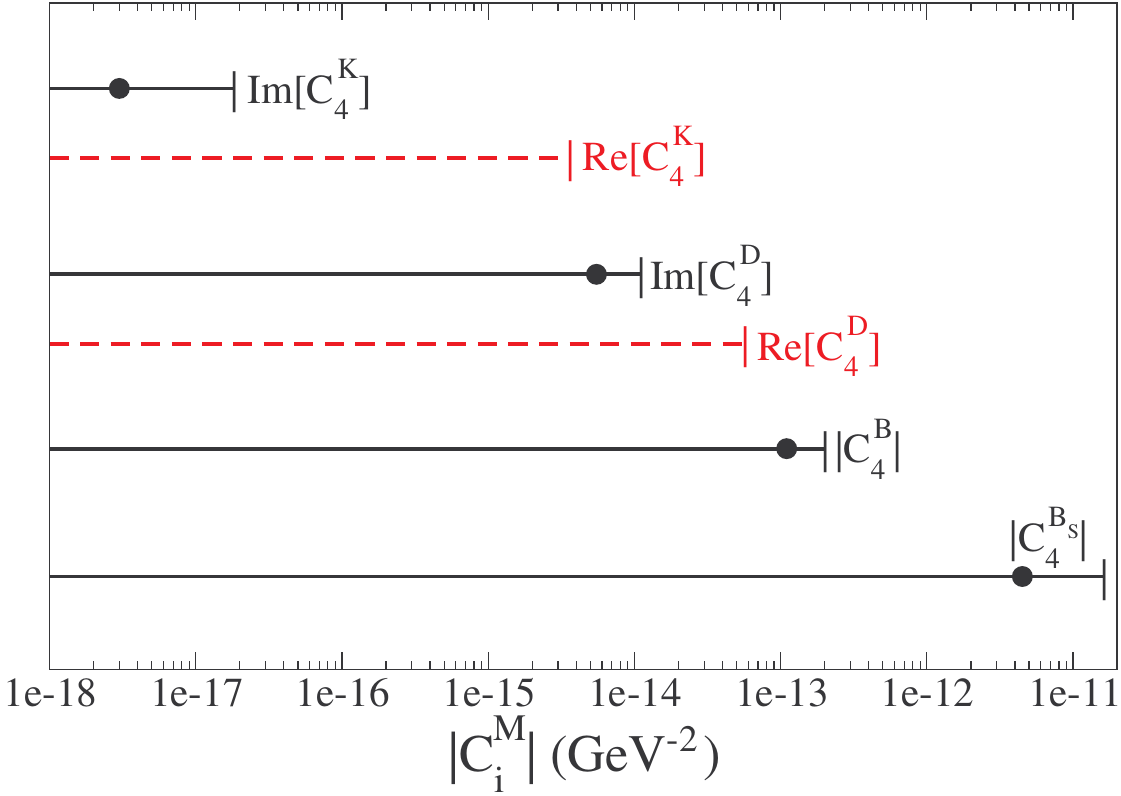}}
\caption{Summary of meson mixing limits and predictions (dots) for the
Wilson coefficients that are most constraining for our model.  Allowed
ranges for Im($C_i^j$) are shown as solid lines, while those for
Re($C_i^j$) are dashed, for the cases where there is a distinction.  The new 
neutral Higgs boson mass is assumed to be
115 GeV.
}
\label{pulls}
\end{figure}

\section{Numerical determination of couplings}
\label{numerical}

We now demonstrate numerically that it is possible to find values of
the parameters consistent with all observables.  We continue to assume
that $m_H \cong m_A$ for the new neutral Higgs bosons, and adopt the 
benchmark choice $m_H =115\,$GeV, 
while taking $m_\pm = 100\,$GeV and $\rhoett=2.5$, consistent with 
a Higgs mixing angle $\cba\cong -8\times 10^{-3}$
from eq.\ (\ref{rhoett_bound}). 
These values also satisfy collider constraints as long as
$\rho_d^{bb},\rho_u^{tt}$ are sufficiently small, as we will verify, and 
 $Z\to\ell\ell$ universality.  

The best-fit values of the quark couplings $\rho_{d}$ are
determined using a $\chi^2$ statistic that incorporates
the most constraining
observables (in addition to the anomalies we set out to address), 
namely the tree-level contributions to meson mixing. We
minimize $\chi^2$ with respect to the elements of $\rho_d$,
with $\rho_u$ determined by the CT ansatz (\ref{CTM}), requiring that
the upper limits on the Wilson coefficients $C_4^M$ not be exceeded
for any meson $M$. Minimizing $\chi^2$
leaves some degeneracy in the fit with respect to products of the
form $\rho_q^{ij}\rho_q^{ji}$, which generally must be small to
satisfy the mixing constraints.  We partially resolve this degeneracy by
trying to enforce $|\rho_q^{ij}|\sim |\rho_q^{ji}|$ as much as
possible,  to avoid having matrix elements that are unnaturally small,
as we will discuss in section \ref{naturalness}.

We make the simplifying approximation of 
real-valued $\rho_d$ and $\rho_u$.\footnote{However we do not assume
that phases are small when applying the limits on Wilson coefficients
from meson mixing, where the bounds on imaginary parts can be orders
of magnitude stronger than on the real parts.  We allow for
the possibility that the
phases are $O(1)$ for the interpretation of these bounds, by
imposing the more stringent imaginary part limits.}
  This requires ignoring the phase
of the CKM matrix as well since 
$\rho_u = \eta V\rho_d^\dagger V^\dagger U$
according to (\ref{CTM}), where we take $\eta=0.8$ and
$U={\rm diag}(-1,1,-1)$ for
definiteness.  We therefore approximate $V$ as an SO(3) matrix using eqs.\ (12.3-12.4)
of \cite{PDG} with the replacement $\bar\rho+i\bar\eta\to (\bar\rho^2+
\bar\eta^2)^{1/2}$, leaving for future work to incorporate phases
into the analysis.  

Using this fitting procedure, 
an example of couplings that are consistent with all constraints is
\bea
\label{rhodfit}
	\rho_d &=& \left(\begin{array}{ccc}
 \phm 8.3\cdot 10^{-4} &\phm 1.1\cdot 10^{-7} & \phm1.3\cdot 10^{-4} \\
  \phm3.8\cdot 10^{-7} & \phm7.7\cdot 10^{-4} & \phm2.8\cdot 10^{-3} \\
 -1.2\cdot 10^{-5} & -2.1\cdot 10^{-5} & \phm5.5\cdot 10^{-2}
	\end{array}\right) 
\eea
\bea
\label{rhoufit}
	\rho_u &=& \left(\begin{array}{ccc}
 -6.6\cdot 10^{-4} & \phm3.5\cdot 10^{-6} &  \phm1.4\cdot 10^{-4} \\
  -2.1\cdot 10^{-5} & \phm7.8\cdot 10^{-4}  &\phm1.8\cdot 10^{-3} \\
  -7.6\cdot 10^{-4} & \phm4.0\cdot 10^{-3}  &\phm4.4\cdot 10^{-2}
	\end{array}\right)
\eea
\bea
	V\rho_d &=& \!\!\!\left(\begin{array}{rrr}
 8.1\cdot 10^{-4} & 1.7\cdot 10^{-4} & \phm9.5\cdot 10^{-4} \\
 -1.9\cdot 10^{-4} & 7.5\cdot 10^{-4}  & \phm5.0\cdot 10^{-3} \\
 -6.8\cdot 10^{-6} & -5.3\cdot 10^{-5}  & \phm5.5\cdot 10^{-2}
	\end{array}\right)
\label{rhofits}
\eea
Recall that only $\rho_d$ is independent; $\rho_u$ is determined,
and the charged Higgs couplings $V\rho_d$ are shown for convenience.
Other solutions can be found with smaller values of the 
matrix elements not needed for the $B$ decay anomalies
($\rho_d^{bb}$ and $\rho_d^{sb}$); 
we have allowed the former to be nearly
as large as is consistent with meson mixing constraints. 

In fig.\ \ref{pulls} we show the predicted values versus experimental
limits on the magnitude of the $C_{4}^{K,D,B,B_s}$ Wilson coefficients
corresponding to the tree-level contributions to meson mixing from $H$
exchange.  For $K^0$ and $D^0$ we satisfy the more stringent constraints
on the imaginary part of $C_{4}$, noting that ${\rm Im}(C_4^D)$ comes from
the phase of $(V\rho_d^\dagger V^\dagger)_{uc}
(V\rho_d^\dagger V^\dagger)_{cu}^*\cong
(V_{ub}V_{cs}^*|\rho_d^{bb}|)^2$, which is 
of the same order as the real part.    

For the values of $\rho_d^{bb}$ and $\rho_d^{sb}$ given 
in (\ref{rhodfit}), the
cross section $\sigma(bbH) = 1.2\,$pb for production of $H$
by $bb$ fusion, not far below the CMS upper limit of $1.8\,$pb, 
while the branching ratio for $B_s\to\ttpm$ is predicted to be
$2.9\%$, close to the current upper limit of $5\%$. 

\subsection{Including $\tau\to K\nu$ deficit}

In the preceding fit we did not try to obtain the negative value of
$(V\rho_d)_{us}$ favored by eq.\ (\ref{Vruseq}) for explaining the
low $\tau\to K\nu$ determination of $V_{us}$.  Doing so 
introduces some tension with the limit on $b\to s\gamma$.  We are
able to obtain $(V\rho_d)_{us} = -1.8\times 10^{-3}$, so that
$(V\rho_d+ \rho_u^\dagger V)_{us} = (1-\eta)(V\rho_d)_{us} = -3.6\times
10^{-4}$, close to the target value of (\ref{Vruseq}), while respecting
all other constraints except for a marginal violation of the $2\,\sigma$
limit on $b\to s\gamma$.  The fit gives $C'_7 = -0.036$, which is
still in the $3\,\sigma$ allowed region of ref.\ 
\cite{Descotes-Genon:2015uva}.

\section{One-loop corrections to couplings}
\label{naturalness}

A texture present in the $\rho$ matrices at tree level gets modified by  loops
involving products of $\rho_{u,d}$ as well as the  CKM matrix $V$. Rather than
estimating all possible loop corrections, it is more efficient to use a spurion
analysis in which the Yukawa matrices are taken to
transform under the full SU(3)$_u\times$SU(3)$_d\times$SU(3)$_Q$
flavor symmetries, constructing all combinations that transform in 
the same way as
the couplings of interest.  This generates a large subset of the complete
set of flavor structures that should arise from the loop corrections.

\begin{figure}[t]
\centerline{
\includegraphics[width=0.9\columnwidth]{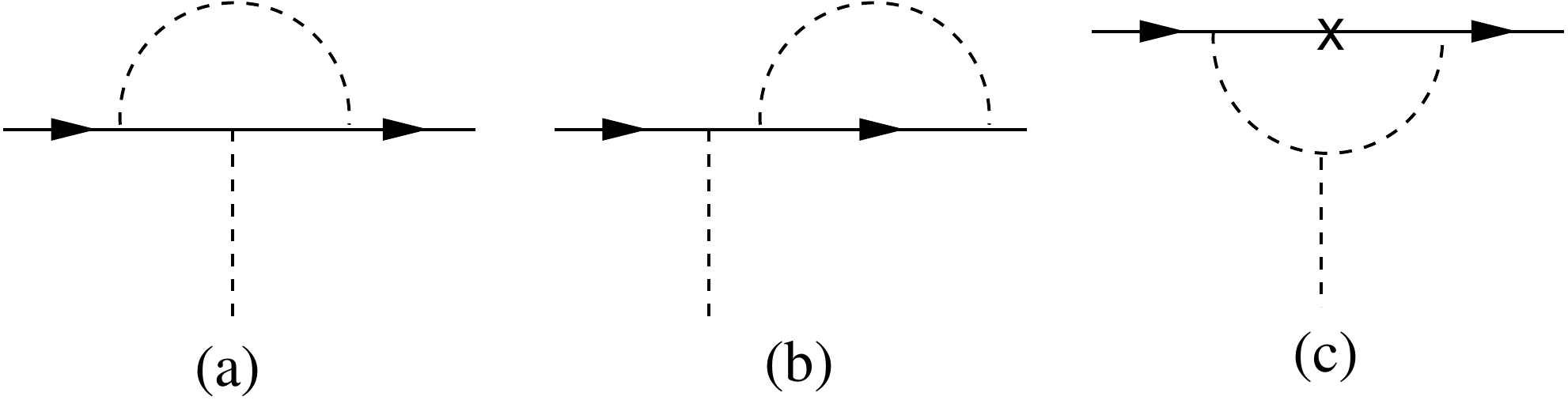}}
\caption{(a,b): One-loop corrections to couplings captured by the
spurion analysis.  (c) Correction requiring a chirality flip, which is not
explicitly included in the spurion analysis.}
\label{spurion}
\end{figure}

The procedure captures the contributions from 
loops carrying momenta between the fundamental scale down to the scale of
electroweak symmetry breaking.  In particular, it accounts for one-loop
diagrams of the type shown in fig.\ \ref{spurion}(a,b).  Diagrams of the
type \ref{spurion}(c) require a mass insertion in the fermion line,
which needs a more detailed computation.  We defer such a study to the future,
hoping that the terms included are reasonably representative of the 
full corrections.  It is also possible that they give an overestimate of the
true corrections, as the example of $Z\to\ttpm$ in section \ref{Zll}
showed.  In that process, the perturbation to the vertex turn out to be 
considerably smaller than a naive estimate of the loop diagrams suggested.

It is clearest to think initially in the unbroken phase, using the
couplings $\hat y_{u,d,e}$ and $\hat\rho_{u,d,e}$ of the Lagrangian 
(\ref{general}) in the original field basis before diagonalizing
$\hat y_{u,d,e}$.  The spurious transformation properties of the 
Yukawa matrices under the flavor symmetries are
\bea
(\hat y_u,\,\hat\rho_u) &\rightarrow& V_Q^{\dagger} 
\, (\hat y_u,\,\hat\rho_u) V_u,\nn\\
(\hat y_d,\,\hat\rho_d) &\rightarrow& V_Q^{\dagger} \, (\hat y_d,\,\hat\rho_d) \, V_d, \nn \\
(\hat y_e,\rho_e) &\rightarrow& V_L^{\dagger} \, (\hat y_e,\rho_e)\, V_e,
\eea
where $V_i$ denotes an element of the SU(3)$_{i}$ 
flavor subgroup.

At one loop, corrections that are cubic in the couplings are generated.
Purely on the basis of the symmetries, we see that the following matrix
structures would be allowed (now considering only the quark couplings):
\bea
	\delta(\hat y_u,\,\hat\rho_u) \sim \left\{{\phm[(\hat y_u,\,\hat\rho_u)\cdot
	(\hat y_u^\dagger,\,\hat\rho_u^\dagger)]\cdot (\hat y_u,\,\hat\rho_u)\atop
	+ [(\hat y_d,\,\hat\rho_d)\cdot (\hat y_d^\dagger,\,\hat\rho_d^\dagger)]\cdot
	(\hat y_u,\,\hat\rho_u)}\right.\nn\\
	\delta(\hat y_d,\,\hat\rho_d) \sim \left\{{\phm[(\hat y_u,\,\hat\rho_u)\cdot
	(\hat y_u^\dagger,\,\hat\rho_u^\dagger)]\cdot (\hat y_d,\,\hat\rho_d)\atop
	+ [(\hat y_d,\,\hat\rho_d)\cdot (\hat y_d^\dagger,\,\hat\rho_d^\dagger)]\cdot
	(\hat y_d,\,\hat\rho_d)}\right.
\eea
Hence flavor symmetry alone allows 16 possible combinations as corrections
to each kind of coupling.  In practice, not all of these are realized by
the diagrams in fig.\ \ref{spurion}, as we will explicitly check.
Moreover, for a coupling to a given external Higgs field,
half of these are suppressed by 
the small mixing angle $\cba$, since for example the product of
the two vertices 
associated with a
loop involving $H$ goes like $\hat\rho^2 + \cba\hat\rho \hat y + \cba^2 \hat y^2$,
while those connected to $h$ give $\hat y^2 + \cba\hat\rho \hat y + \cba^2 \hat\rho^2$.
Finally, it is convenient once the appropriate structures are identified
to transform to the basis where the fermion mass matrices are
diagonalized, and express the results in terms of $y_i$ (the diagonalized
version of $\hat y_i$) and $\rho_i$.  
This introduces factors of the CKM matrix $V$ wherever there
is a mismatch between $u$- and $d$-type indices.  These come from diagrams
with charged Higgs exchange.

\subsection{Corrections to quark couplings}

Beyond tree level, we can no longer characterize the nonstandard couplings
by just $\rho_d$ and $\rho_u$ because the simple relation between the 
nonstandard couplings of the light Higgs $h$ and the couplings of the heavy
Higgs bosons is not preserved.  The nonstandard Yukawa couplings get corrections of the
form
\bea
	\delta{\cal L}_q &=& 
	-\sum_{q=u,d}\bar q_L \left({h\over\sqrt{2}}\,\delta y_{q,h}
	+ H\delta\rho_{q,H}\right)q_R 
	\\
	&-& H^+ \bar u\left(\delta(V\rho_d)_\pm P_{\sss R} -  
	\delta(\rho^\dagger_u V)_\pm P_{\sss L}\right)d +{\rm h.c.} \nonumber
\eea
where $H = (H^0+iA^0)/\sqrt{2}$.  From examination of the diagrams in 
fig.\ \ref{spurion}(a,b), we estimate the corrections
\bea
\delta y_{u,h} &=& \epsilon_u^1\,y_u y_u^\dagger y_u  +  
\epsilon_u^2\, \rho_u\rho_u^\dagger y_u 
+ \epsilon_u^3\, y_u\rho_u^\dagger \rho_u \nn \\
&+& \epsilon_u^4 \, \rho_u y^\dagger_u \rho_u  +  
\epsilon_u^5 \,V\rho_d y_d^\dagger V^\dagger\rho_u 
+ \epsilon_u^6\, V\rho_d\rho_d^\dagger V^\dagger y_u\nn \\
&+& \cba\{\epsilon_u^7\,\rho_u\rho_u^\dagger\rho_u + 
\epsilon_u^8\, y_u\rho_u^\dagger y_u + \epsilon_u^9\, \rho_u y_u^\dagger y_u\
\nn\\
&+& \epsilon_u^{10}\, y_u y_u^\dagger\rho_u + 
\epsilon_u^{11}\, V\rho_d\rho_d^\dagger V^\dagger\rho_u\}
\eea
\bea
\delta y_{d,h}&=& 
\epsilon_d^1\,y_d y_d^\dagger y_d  +  
\epsilon_d^2\, \rho_d\rho_d^\dagger y_d 
+ \epsilon_d^3\, y_d\rho_d^\dagger \rho_d \nn \\
&+& \epsilon_d^4 \, \rho_d y^\dagger_d \rho_d  +  
\epsilon_d^5 \,V^\dagger\rho_u y_u^\dagger V\rho_d 
+ \epsilon_d^6\, V^\dagger\rho_u\rho_u^\dagger V y_d\nn \\
&+& \cba\{\epsilon_d^7\,\rho_d\rho_d^\dagger\rho_d + 
\epsilon_d^8\, y_d\rho_d^\dagger y_d + \epsilon_d^9\, \rho_d y_d^\dagger y_d\
\nn\\
&+& \epsilon_d^{10}\, y_d y_d^\dagger\rho_d + 
\epsilon_d^{11}\, V^\dagger\rho_u\rho_u^\dagger V\rho_d\}
\eea
\bea
\delta y_{u,H} &=& \cba\{\eta_u^1\,y_u y_u^\dagger y_u  +  
\eta_u^2\, \rho_u\rho_u^\dagger y_u 
+ \eta_u^3\, y_u\rho_u^\dagger \rho_u \nn \\
&+& \eta_u^4 \, \rho_u y^\dagger_u \rho_u  +  
\eta_u^5 \,V\rho_d y_d^\dagger V^\dagger\rho_u 
+ \eta_u^6\, V\rho_d\rho_d^\dagger V^\dagger y_u\}\nn \\
&+& \eta_u^7\,\rho_u\rho_u^\dagger\rho_u + 
\eta_u^8\, y_u\rho_u^\dagger y_u + \eta_u^9\, \rho_u y_u^\dagger y_u\
\nn\\
&+& \eta_u^{10}\, y_u y_u^\dagger\rho_u + 
\eta_u^{11}\, V\rho_d\rho_d^\dagger V^\dagger\rho_u
\eea
\bea
\delta y_{d,H}&=& 
\cba\{\eta_d^1\,y_d y_d^\dagger y_d  +  
\eta_d^2\, \rho_d\rho_d^\dagger y_d 
+ \eta_d^3\, y_d\rho_d^\dagger \rho_d \nn \\
&+& \eta_d^4 \, \rho_d y^\dagger_d \rho_d  +  
\eta_d^5 \,V^\dagger\rho_u y_u^\dagger V\rho_d 
+ \eta_d^6\, V^\dagger\rho_u\rho_u^\dagger V y_d\}\nn \\
&+& \eta_d^7\,\rho_d\rho_d^\dagger\rho_d + 
\eta_d^8\, y_d\rho_d^\dagger y_d + \eta_d^9\, \rho_d y_d^\dagger y_d\
\nn\\
&+& \eta_d^{10}\, y_d y_d^\dagger\rho_d + 
\eta_d^{11}\, V^\dagger\rho_u\rho_u^\dagger V\rho_d
\eea
\bea
\delta(V\rho_d)_\pm  &=&  \zeta_d^1\,\rho_u\rho_u^\dagger V\rho_d + 
	\zeta_d^2\, y_u\rho_u^\dagger V y_d + 
	\zeta_d^3\, y_u y_u^\dagger V\rho_d\nn\\
	&+& \zeta_d^4\, V\rho_d\rho_d^\dagger\rho_d + 
	\zeta_d^5\, V\rho_d y_d^\dagger y_d\nn\\
	&+& \cba\{ \zeta_d^6\, y_u\rho_u^\dagger V\rho_d
	+ \zeta_d^7\, \rho_u\rho_u^\dagger V y_d 
	+ \zeta_d^8\, \rho_u y_u^\dagger V\rho_d\nn\\
	&+& \zeta_d^9\, V\rho_d\rho_d^\dagger y_d
	+ \zeta_d^{10}\, V\rho_d y_d^\dagger \rho_d\}
\eea
\bea
\delta(\rho_u^\dagger V)_\pm  &=&  
	\zeta_u^1\,\rho_u^\dagger V\rho_d\rho_d^\dagger
	+\zeta_u^2\, y_u^\dagger V\rho_d y_d^\dagger 
	+\zeta_u^3\, \rho_u^\dagger V y_d y_d^\dagger \nn\\
	&+& \zeta_u^4\, \rho_u^\dagger\rho_u\rho_u^\dagger V+ 
	\zeta_u^5\, y_u^\dagger y_u\rho_u^\dagger V \nn\\
	&+& \cba\{ \zeta_u^6\, \rho_u^\dagger V\rho_d y_d^\dagger
	+ \zeta_u^7\, y_u^\dagger V \rho_d\rho_d^\dagger 
	+ \zeta_u^8\, \rho_u^\dagger V y_d \rho_d^\dagger \nn\\
	&+& \zeta_u^9\, y_u^\dagger \rho_u\rho_u^\dagger V
	+ \zeta_u^{10}\, \rho_u^\dagger y_u \rho_u^\dagger V \}
\eea
Here the coefficients $\epsilon_f^{i},\,\eta_f^i,\,\zeta_f^i$  are all 
assumed to be of order 
$1/16 \pi^2$.  The contributions that are suppressed by 
$\cba$ can be understood as having an odd 
or even number of 
$\rho$ or $y$ insertions respectively.  For completeness, 
we include two corrections that exist
also within the standard model, namely $\epsilon_{u,d}^{1}$.  We omit them
from the following analysis since they do not involve the new physics we
are investigating.

To test the degree of tuning required by our numerical fit, we have computed
the maximum of each of these estimates using the numerical values of
the couplings in 
(\ref{rhofits}).  The magnitude of correction to each coupling, relative to 
its tree-level value, and the correction responsible for the largest
effect in each matrix, is given by
\be
\label{drhod}
	\left|{\delta\rho_{d,H}\over\rho_d}\right| = 
	\left(\begin{array}{ccc}
 10^{-8} &   10^{-6} &   10^{-4} \\
 10^{-5} &   10^{-7} &   10^{-5} \\
 10^{-5} &   10^{-4} &   10^{-5}
	\end{array}\right),\quad \eta_d^8
\ee
\be
\label{drhou}
	\left|{\delta\rho_{u,H}\over\rho_u}\right| = 
	\left(\begin{array}{ccc}
 10^{-7} &   10^{-4} &   10^{-2} \\
 10^{-4} &   10^{-5} &   10^{-2} \\
 10^{-2} &   10^{-2} &   10^{-2} 
	\end{array}\right),\quad
\!\eta_u^{8\xhyphen 10}\!\!\!\!\!\!\!
\ee
\be
\label{dydh}
	\left|{\delta y_{d,h}\over\cba\,\rho_d}\right| = 
	\left(\begin{array}{ccc}
 10^{-5} &   0.2 &   0.2 \\
 0.01 &   10^{-4} & 10^{-2} \\
 0.1 & 0.6 &   0.2
	\end{array}\right),\quad \epsilon_d^5
\ee
\be
\label{dyuh}
	\left|{\delta y_{u,h}\over\cba\,\rho_u}\right| = 
	\left(\begin{array}{ccc}
 10^{-3} &   0.9 &   2 \\
 0.3 &   0.05 & 0.8 \\
 0.2 & 0.2 &   0.4
	\end{array}\right),\quad \epsilon_u^6
\ee
\be
\label{drhom}
	\left|{\delta(V\rho_d)_\pm\over V\rho_d}\right| = 
	\left(\begin{array}{ccc}
 10^{-9} &   10^{-7} &   10^{-5} \\
 10^{-7} &   10^{-7} & 10^{-5} \\
 10^{-2} & 10^{-2} &   10^{-2}
	\end{array}\right),\quad \zeta_d^3
\ee
\be
\label{drhop}
	\left|{\delta(\rho_u^\dagger V)_\pm\over \rho_u^\dagger V}\right| = 
	\left(\begin{array}{ccc}
 10^{-7} &   10^{-5} &   10^{-5} \\
 10^{-6} &   10^{-5} & 10^{-5} \\
 10^{-2} & 10^{-2} &   10^{-2}
	\end{array}\right),\quad \zeta_u^5
\ee

The most potentially worrisome elements are the corrections to
$y_{d,h}^{bs}$ and $y_{u,h}^{uc}$, which can increase the tree-level
contributions to $D$ and $B_s$ mixing mediated by light Higgs
exchange.  The relatively large corrections to $y_{u,h}$,  
namely $\delta
y_{u,h}^{ut},\delta y_{u,h}^{ct}\sim 1$, are harmless since they only
affect  flavor-changing decays of the top quark, which are weakly
 constrained by observations.  The other corrections can
perturb the predictions for the $C_2$ mixing coefficients in 
eq.\ (\ref{C2coeffs}) by factors of at most $O(1)$.  But these
coefficients are less constraining than the $C_4$'s in our fit.  
The one that comes closest is $C_2^{B_s}$ which is $0.06$ of the 
experimental limit.  Thus there is plenty of room for the tree-level
couplings to receive corrections of the order we find without 
violating any experimental constraints.

\subsection{Lepton couplings}

Unlike for the quark couplings, naturalness does not require us
to turn on any significant off-diagonal elements in $\rho_e^{ij}$.
In the absence of neutrino masses, these are not generated by
loops.  Charged Higgs exchange generates an off-diagonal coupling
of order
\be
	\delta y_{e,h}^{\mu\tau} \sim {\lambda_1 \rhoett
	\over 16\pi^2}\, {m_\mu m_\nu\over m_\pm^2}
\ee
which is negligible.  This conclusion would also remain true if we allowed
for nonvanishing $\rho_e^{ee}$ and $\rho_e^{\mu\mu}$ entries (with
smaller values than $\rhoett$).  We do not pursue a more
complete exploration of the allowed leptonic couplings here.

\subsection{Higgs potential coupings}
We can estimate the size of corrections to the Higgs potential
couplings $\lambda_i$ more definitely than those for the quark 
couplings since the beta functions are known; see for example
ref.\ \cite{Cline:2011mm}.  Our scenario requires that $\lambda_3\ll 1$
and $\lambda_5\ll 1$, whereas the other $\lambda_i$ could be larger.
Taking $\epsilon = 1/16\pi^2$ (which ignores possible logarithmic
enhancements), the
dominant contributions to the one-loop corrections are of order
\bea
\label{dlambda}
\delta\lambda_1 &\sim& \epsilon\left[
	(\lambda+\lambda_6)(6\lambda_1+2\lambda_2) + 2\lambda_1^2
	+ \lambda_2^2 + 2\lambda_4^2\right.\nn\\ 
        && \quad+\left.\lambda_1(\rho_e^{\tau\tau})^2+3y_t^2-\sfrac92 g^2)
	+\sfrac98 g^4 - 6 y_t^2(\rho_u^{tt})^2\right]\nn\\
	&&\sim 3\cdot 10^{-2}\nn\\
\delta\lambda_2 &\sim&
	\epsilon\left[2\lambda_2(\lambda+\lambda_6) +
	4\lambda_1\lambda_2 + 2\lambda_2^2 + 5\lambda_4^2 \right.\nn\\
	&& \quad+\left.\lambda_2(\rho_e^{\tau\tau})^2+3y_t^2-\sfrac92 g^2) - 6y_t^2(\rho_u^{tt})^2
	\right]\nn\\ && \sim 9\cdot 10^{-3}\nn\\
\delta\lambda_3 &\sim& \epsilon\left[
	\sfrac52\lambda_4^2 - 3 y_t^2(\rho_u^{tt})^2\right]
	\sim 1\cdot 10^{-3}\nn\\
\delta\lambda_4 &\sim& \epsilon\left[ \lambda_4 (12\lambda_6
        +3\lambda_1 + 4\lambda_2 -\sfrac92 g^2 +\sfrac32 y_t^2 + 
	\sfrac12(\rho_e^{\tau\tau})^2)\right.
	\nn\\
      && \left.\quad -6y_t (\rho_u^{tt})^3\right]\quad \sim 2\cdot 10^{-2}\nn\\
\delta\lambda_5 &\sim& \epsilon\left[
	\lambda_4(3\lambda_1+ 2\lambda_2) - 
	6 y_t^3\rho_u^{tt}\right]\sim 4\cdot 10^{-4}\nn\\
\delta\lambda_6 &\sim& \epsilon\left[ 12\lambda_6^2 
	+ \lambda_1^2 + \lambda_1\lambda_2 +\sfrac12\lambda_2^2
	+ 6\lambda_4^2 \right.\nn\\
        && \quad+\left. \sfrac{9}{16}g^4 - \sfrac92\lambda_6 g^2
	-6(\rho_u^{tt})^4 - 2(\rho_e^{\tau\tau})^4\right]\nn\\
	&& \sim -0.5
\eea
where $g$ is the SU(2)$_L$ gauge coupling and we have ignored terms
involving $g'$ (the SU(1) hypercharge) and the small $\lambda_3$ and
$\lambda_5$ couplings. We have included the effect of 
$\rho_e^{\tau\tau}$ where it is not suppressed by powers of the SM
tau Yukawa coupling.
To obtain the numerical estimates, we chose
fiducial values of the other couplings that are consistent with
the assumed mass spectrum $m_\pm=100{\rm\,GeV}$, $m_H=115{\rm\,GeV}$,
\be
\label{fiducial}
	\lambda_1 = 0.3,\quad \lambda_2 = 0.1,\quad \lambda_4=0.3,
	\quad \lambda_6 = 0.7
\ee

The potentially worrisome corrections are those for the smallest
couplings, $\lambda_5\cong -6\times 10^{-4}$ using $\cba=-8\times
10^{-3}$ (see eq.\ (\ref{rhoett_bound})) and eq.\ (\ref{mixing}), and
$\lambda_3 \cong 10^{-3}$.  Comparison with the estimates in 
(\ref{dlambda}) indicates that these values are relatively stable. Our
choice of couplings in (\ref{fiducial}) allows for some accidental
cancellation in $\delta\lambda_5$ between the bosonic and fermionic
loops.  Even without such a cancellation, the contribution from the
top quark by itself is $\sim 3\lambda_5$ which requires only a mild 
coincidence between tree and loop contributions to obtain the desired
value.  Although the correction to $\lambda_6$ is relatively large, the phenomenology
of the model is largely insensitive to its value.

\subsection{Landau pole}
The large coupling $\rho_e^{\tau\tau}\cong 2.5$ may be expected to give
rise to a Landau pole at a relatively low scale, indicating that further
new physics will be required to achieve a UV complete description.  To
estimate this scale we consider the renormalization group equations that
depend most sensitively on $\rho_e^{\tau\tau}$:
\bea
	{d\lambda_6\over d\ln\mu^2} &\cong& {1\over 16\pi^2}
	\left(12\lambda_6 -
2(\rho_e^{\tau\tau})^4\right),\nn\\
	{d\rho_e^{\tau\tau}\over d\ln\mu^2} 
	&\cong& {1\over 16\pi^2} (\rho_e^{\tau\tau})^3
\eea
Numerically solving using the initial conditions $\lambda_6=0.7$
(see eq.\ (\ref{fiducial})) and $\rho_e^{\tau\tau}\cong 2.5$ at the
scale $\mu=100\,$GeV, we find that the couplings diverge at $\mu\cong
55\,$TeV.

\begin{figure}[t]
\centerline{
\includegraphics[width=0.9\columnwidth]{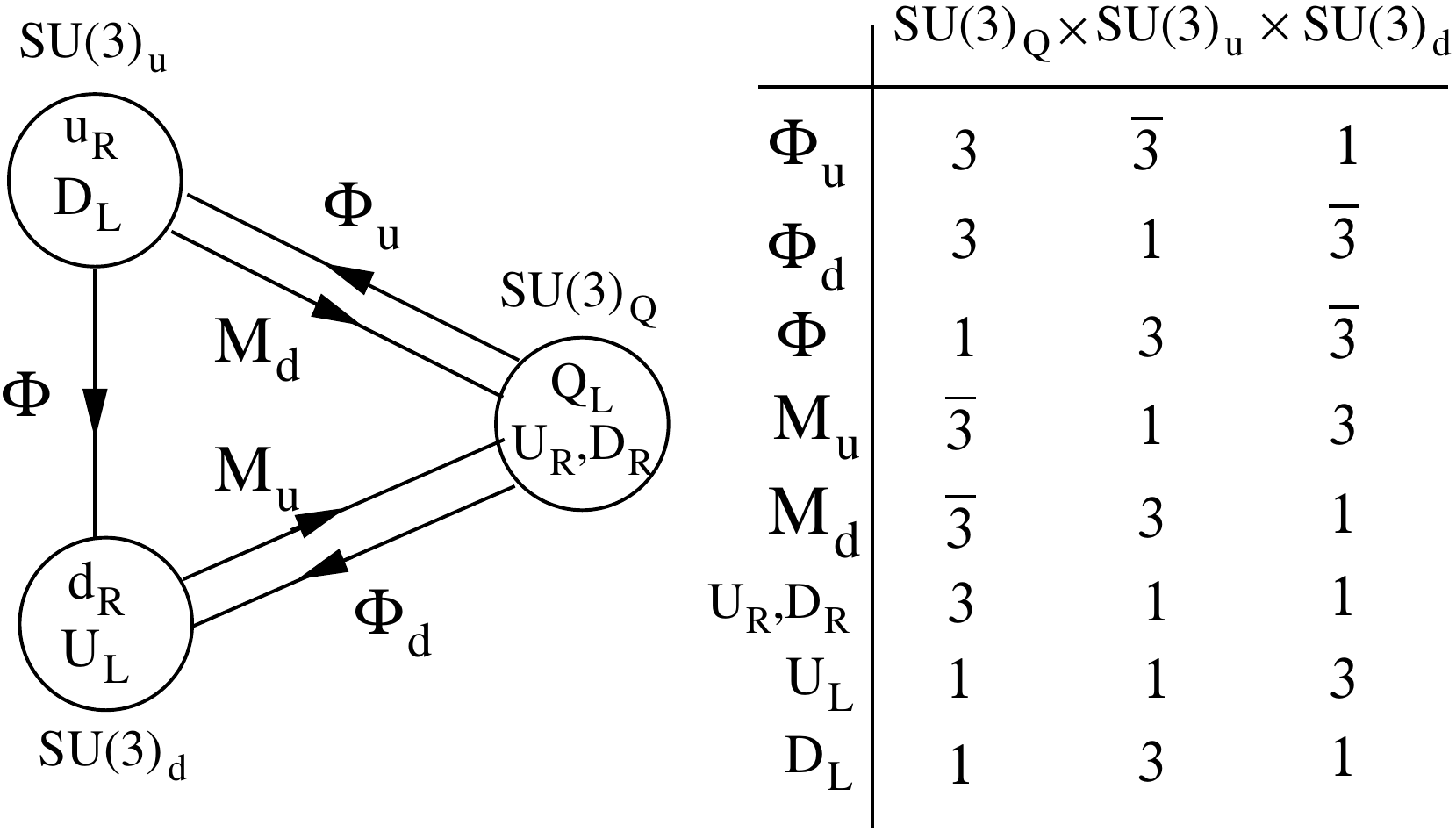}}
\caption{Moose diagram indicating the transformation properties of the
SM quark fields $u_R$, $d_R$, $Q_L$, the heavy singlet quarks $U,D$, and the
bifundamental $\Phi,\Phi_u,\Phi_d,M_u,M_d$ under the 
SU(3)$_Q\times$SU(3)$_u\times$SU(3)$_d$ flavor symmetry .}
\label{moose}
\end{figure}

\section{Microscopic origin of CT ansatz}
\label{uvmodel}

\begin{table}[b]
\begin{tabular}{|c|c|c|c|c|c|c|c|c|}
\hline
$H_1$ & $H_2$ & $\Phi$ & $\Phi_{u,d}$& $M_{u,d}$ & $Q_L$ & $(u_R,d_R)$ & 
	$(U_L,D_L)$ & $(U_R,D_R)$ \\
\hline
$1$ & $-1$ & $1$ & $1$ & $-1$ & $i$ & $i$ & $i$ & $-i$\\
\hline
\end{tabular}
\caption{$Z_4$ charge assignments needed for the allowed terms in the
Lagrangian (\ref{UVL}).}
\label{tab:charges}
\end{table}

As an example of what kind of physics could give rise to the CT ansatz
(\ref{CTM}), we construct a model where the 
SU(3)$_Q\times$SU(3)$_u\times$SU(3)$_d$ flavor symmetry is spontaneously 
broken by bifundamentals $\Phi_u,\Phi_d,\Phi,M_u,M_d$, coupling to heavy
SU(2)-singlet quarks $U_{R,L}$ and $D_{R,L}$.  The charges of the fields
under the flavor symmetries are shown in fig.\ \ref{moose}.  As in 
(\ref{general}), $H_1$ is the SM-like Higgs field and $H_2$ is the new doublet,
before mixing of the neutral mass eigenstates, and we take the Lagrangian at 
the high scale to be
\bea
	{\cal L} &=& {1\over \Lambda}\left(H_1 \bar Q_L \Phi_u u_R + \tilde H_1 \bar Q_L \Phi_d
d_R\right)\nonumber\\
	&+&\qquad\!\! H_2 \bar Q_L U_R \ \quad +  \tilde H_2 \bar Q_L D_R\nonumber\\
	&+& \qquad\!\! \bar U_L \Phi^\dagger u_R \qquad \!\!+ \bar D_L \Phi d_R\nonumber\\
	&+& \qquad\!\! \bar U_L M_u U_R \qquad \!\!\!\!+ \bar D_L M_d D_R
\label{UVL}
\eea
which respects the full flavor symmetry.  In table \ref{tab:charges}
we show the charge assignments under a $Z_4$ symmetry that allows the 
interactions in (\ref{UVL}) while forbidding those with $H_1$ and $H_2$
interchanged.  This symmetry gets spontaneously broken by VEVs of the
bifundamental fields $M_{u,d}$, allowing for subsequent generation of the terms in 
the Higgs potential (\ref{pot}) that break the symmetry ({\it i.e}., the
terms with coefficients $\lambda_4$ and $\lambda_5$).  The $m_{12}^2 H_1^\dagger
H_2$ term that breaks it softly can be allowed from the outset, to avoid
cosmological problems from domain walls. 

In (\ref{UVL}) we have not specified
a fully renormalizable Lagrangian, but merely assumed that the SM-like
Yukawa couplings arise from the VEVs, $\hat y_{u,d} = \langle \Phi_{u,d}\rangle/\Lambda$
with some large mass scale $\Lambda$.  Our main interest is in the origin of
the new Yukawa couplings $\hat\rho_{u,d}$.  Assuming the simple symmetry-breaking
pattern $\langle M_d\rangle = \langle M_u\rangle/\eta = M$ times the unit matrix
in flavor space, after integrating out the heavy $U,D$ quarks the new
Yukawas are given by $\hat\rho_{u} = \eta\langle\Phi^\dagger\rangle/M$,
$\hat\rho_d = \langle\Phi\rangle/M$.  However this is in the basis where $\hat y_{u,d}$
are not yet diagonalized.  As usual, we must transform $u_R \to R_u^\dagger u_R$, 
$d_R \to R_d^\dagger  d_R$, $u_L\to L_u^\dagger u_L$, 
$d_L\to L_d^\dagger d_L$.  In the
quark mass basis, $\rho_u = \eta L_u \langle\Phi^\dagger\rangle R_u^\dagger/M$,
$\rho_d = L_d \langle\Phi\rangle R_d^\dagger/M$.

With these results, we can now explain the origin of the ansatz (\ref{CTM}) by computing the two sides of that
relation:
\bea
	\eta\,UV\rho_d &=& \eta\,UL_u {\langle\Phi\rangle\over M} R_d^\dagger\nonumber\\
	\rho_u^\dagger V &=& \eta R_u {\langle\Phi\rangle\over M} L_d^\dagger
\eea
where we have used $V = L_u L_d^\dagger$. 
Equality of $\eta\, U V\rho_d$ and $\rho^\dagger_u V$ follows from taking
\be
	R_u = U L_u,\quad R_d = L_d
\ee
One recognizes the condition $R_d = L_d$ as that which would arise if
$\hat y_d$ is a symmetric matrix.  The other relation $R_u = U L_u$ implies that
$\hat y\, U$ is symmetric.  This means that $\hat y_u$ splits into two pieces,
one symmetric and the other antisymmetric, having no nonvanishing elements in common.
For example if $U={\rm diag}(-1,1,-1)$ then $\hat y_u$ has the structure
\be
\label{hatyu}
	\hat y_u = \left(\begin{array}{ccc}a & 0 & b\\
	0 & c & 0 \\ b & 0 & d \end{array}\right) + 
	 \left(\begin{array}{ccc}0 & e & 0 \\
	-e & 0 & f \\ 0 & -f & 0 \end{array}\right) 
\ee 
We imagine that it is possible to find a potential $V(\Phi_u)$ whose minimum
has this form.   Then our
ansatz, which at first sight appears contrived, can be a simple consequence
of the SM-like Yukawa matrix $\hat y_d$ being symmetric in the underlying theory
of flavor, while $\hat y_u$ has the pattern (\ref{hatyu}),
along with the ``charge transformation'' bifundamental $\Phi$ whose
VEV gives rise to both $\rho_u$ and $\rho_d$ simultaneously.

Our focus has been to explain the initially peculiar-looking relation
between quark couplings $\rho_u$ and $\rho_d$ proposed herein, rather than
the leptonic couplings $\rho_e$.  From the flavor perspective the ansatz
that $\rho_e$ is dominated by the single element $\rho_e^{\tau\tau}$ is
natural since no dangerous FCNCs arise even for large values of 
$\rho_e^{\tau\tau}$.  However it may be possible to accommodate $\rho_e$
into the UV model in the obvious manner in parallel to the quark
couplings, by adding fields with interactions
\be
	H_2\bar L_L E_R + \bar E_L\Phi E_R + \bar E_L M_e E_R
\ee
We then predict that 
\be
	\rho_e = {\langle\Phi\rangle \over M_e}
\ee
assuming that the mass matrix $M_e$ of the heavy vector-like leptons is
proportional to the unit matrix.  If $\langle\Phi\rangle$ is strongly
dominated by the 33 element, to explain the hypothesized leptonic
couplings, this would lead to quark couplings that are also dominated by
the 33 elements, and with other elements generated from these by
mixing with the approximate structure of the CKM matrix.  Although we do
not pursue this quantitatively here, the values given in (\ref{rhodfit},
\ref{rhoufit}) appear to be roughly consistent with this expectation.

\section{Alternative model}
\label{neutrino}

It is possible to design a similar model that is less constrained by collider
searches,  if a light sterile neutrino $\nu_s$ exists.  The $R(D)$ anomaly
could then be explained by the new process $B\to D^{(*)}\tau\nu_s$
contributing to the observed decays.  Similarly $B$ would get the new decay
channel $B\to\tau\nu_s$, and $H^\pm\to\tau\nu_s$ could contaminate the
$W\to\tau\nu$ signal at LEP.  However the apparent rate for $\tau\to K\nu$
could only be increased in this model because of the lack of interference
with the SM amplitude.  This same absence would also change the fits to the
Wilson coefficients for explaining $R(D)$: we estimate that
\be
  (C^{cb}_{S_R},C^{cb}_{S_L}) = (2.14,\, -1.41)\left({\Lambda\over
	{\rm TeV}}\right)^2
\label{BDfits1}
\ee
by fitting to the decay rates, which are larger than
(\ref{BDfits2a}-\ref{BDfits2b}) to compensate for the lack of interference 
(see appendix \ref{appA}).
It would require a dedicated analysis to check whether this choice of
coefficients 
significantly degrades the agreement with
the decay spectra.

This scenario has the advantage that the stringent collider constraints from
searches for $H\to\tau^+\tau^-$ are evaded, since now 
$H$ decays almost exclusively
into $\nu_s\nu_\tau$. In appendix \ref{appA} we estimate that $m_H$ can
become as large as $175{\rm\, GeV}$, although it is still preferable to
keep $m_\pm$ close to $100{\rm\,GeV}$ to keep the new quark couplings small
so that the predicted branching ratio for
$B\to\nu_s\nu_\tau$ remains reasonably small.  Even though this decay mode
is expected to be less constrained than that for $B\to\tau^+\tau^-$, 
a theoretical understanding
of the total width for $B$ in the SM compared to the experimental value limits
how large it can be.   
We give further details about this alternative model in the appendix.

\section{Conclusions}
\label{conclusion}

The model we have presented is admittedly unlikely, requiring 
coincidences of several new particles and decay modes that are just
below the threshold of detection.  It is much more likely that some of
the experimental anomalies that motivated the model will disappear.  However,
if the $R(D)$ anomaly proves to be real and needs both Wilson coefficients
$C^{cb}_{S_R},\, C^{cb}_{S_L}$ as indicated by several fits to the data, then
our scenario seems to be the only kind of two Higgs doublet model that can be 
compatible with the observations.  The simultaneous explanation of the
other tentative anomalies in $B\to\tau\nu$, $W\to\tau\nu$ (and possibly
$\tau\to K\nu$) is an added bonus that requires little extra model-building
input.

The most striking prediction is that new Higgs bosons of mass  $\sim
100\,$-$\,125{\rm\,GeV}$ that may have been just beyond the kinematic reach of LEP, 
have couplings to $b$ quarks that put the neutral one just below the current
sensitivity of CMS searches. We expect that more data should
soon reveal the existence of the neutral $H$ in the $\tau^+\tau^-$ channel at
the LHC.  A further prediction is that $B_s\to\tau^+\tau^-$ will be
observed with a surprisingly high branching ratio of several percent. The
coupling of the SM-like Higgs boson to $\tau$ should be smaller than the SM
expectation, possibly having the wrong sign. Higher
precision tests of $Z\to\ell\ell$ universality should start to reveal an
excess in $Z\to\ttpm$, and in the invisible $Z$ width due to extra
$Z\to\nu_\tau\bar\nu_\tau$ decays.

The framework also suggests that other observables could be on the edge
of revealing new physics: $\tau\to\pi\nu$, $b\to s\gamma$, and the neutral
meson mixing amplitudes.  These are less definite predictions, since 
we have allowed the new flavor-violating Yukawa couplings $\rho_{u,d}^{ij}$
(apart from those directly involved in explaining $R(D)$) to be nearly as
large as possible while remaining consistent with experimental constraints.
It is possible that they are smaller, even though there is no
fundamental reason that they should be.  By studying the expected size of 
loop contributions to the new couplings, we found that they could indeed be
smaller in many cases without requiring any fine tuning.

Even if all the hints of new physics that motivated this study should 
disappear, some of the ideas presented here could still be of value. First,
the flavor ansatz (\ref{CTM}) reduces the arbitrariness of the new couplings,
allowing us to parametrize everything in terms of $\rho_d$ alone.  In the
absence of anomalous $R(D)$, the need for sizable $\rho_d^{sb}$ would 
disappear and make a symmetric ansatz for $\rho_d^{ij}$ possible, further
reducing the number of independent new Yukawa couplings.  The primary
motivation for our ansatz was to give a more definite flavor structure to
the charged Higgs couplings than is generic if $\rho_u$ and $\rho_d$
are independent.

Second, we have shown that it need not be a disaster to allow  generic new
Yukawa couplings in two Higgs doublet models, even in the absence of any
particular mechanism for suppressing FCNCs.  It could be that the dangerous
couplings are simply small, even though there is no symmetry principle to
explain their smallness. Our model presents a counterexample to the usual
concern, that small values require fine tuning.  We estimated that the relative
corrections from loops to the Yukawa couplings of the new Higgs fields are all 
less than $10^{-2}$, eqs.\ (\ref{drhod}-\ref{drhou},\ref{drhom}-\ref{drhop}).  The
corresponding corrections to the nonstandard couplings of the SM-like Higgs
(\ref{dydh}-\ref{dyuh}) can in some cases exceed their tree-level
values, but this does not lead to any significant FCNCs from $h$ exchange, since
they are still  small enough to remain well below constraints from meson 
oscillations, as long as the Higgs potential coupling $\lambda_3$ that
controls the splitting between $m_H$ and $m_A$ is $\lesssim 10^{-3}$. 

A variant model where the charged Higgs couples to $\bar\nu_s\tau_L$
(where $\nu_s$ is a light sterile neutrino) instead of $\bar\tau_R\nu_\tau$ is outlined in
section \ref{neutrino} and appendix \ref{appA}.  It has greater freedom in the allowed
boson masses and couplings to quarks, making it harder to rule out.  Whether
it can provide as good a fit to $R(D)$ requires further study.

Our original intent was to use a large $\rho_e^{\mu\tau}$ coupling instead of
$\rhoett$ to explain the hint of $h\to\mu^\pm\tau^\mp$ decays of the SM
Higgs seen by CMS and ATLAS
\cite{Khachatryan:2015kon,Aad:2015gha}. This turns out to be much more difficult because of the
$H\to\mu^\pm\tau^\mp$  decay (with 100\% branching ratio) of  the new neutral
boson.  Even though no formal limits on $m_H$ with this  decay channel have
been published, we believe it would have been seen in the searches for
$h\to\mu\tau$ of the SM Higgs, ruling out this model.  Hence a further
prediction of the present model is that $h\to\mu\tau$  events will 
prove to be a statistical fluctuation.

\bigskip
{\bf Acknowledgements.} 
I thank M.\ Trott for inspiring and initially collaborating on this
project, 
  N.\ Arkani-Hamed, B.\ Battacharya, G.\ D'Ambrosio, G.\ Dupuis, A.\ Ferrari, T.\ Gershon, 
J.\ Griffiths,  K.\ Kainulainen,
 Z.\ Ligeti, D.\ London, A.\ McCarn, S.\ Robertson, 
R.\ Sato, S.\ Sekula,  W.\ Shepherd, J.\ Shigemitsu
and  B.\ Vachon for useful
correspondence or discussions, and Mila Huskat for encouragement.
I am grateful to the NBIA for its generous hospitality during this
work, which is also supported by NSERC (Canada).

\begin{appendix}
\section{Sterile neutrino model}
\label{appA}
Here we provide some details relating to the alternative model with the
primary leptonic coupling of the new Higgs fields being to the left-handed
lepton doublets and sterile neutrinos $\nu_R$,
\bea\label{general2}
{\cal L}_Y \ni  
&-& { \bar L}_{L} \, \hat y_\nu \,\tilde H_1 \,  {\nu}_{R} -
{ \bar L}_{L} \,\hat\rho_\nu \, \tilde H_2 \,  {\nu}_{R} 
 + {\rm h.c.} \nn
\eea

There is also a Majorana mass term for
the sterile neutrinos, 
\be \sfrac12(\bar\nu_R M_R \nu_R^c + 
\bar\nu_R^c M_R^* \bar\nu_R)
\ee
Without loss of generality, we can work in the basis where $M_R$ is 
diagonal,
\be
	M_R = {\rm diag}(M_1,\, M_2,\, m_s)
\ee
giving a sterile neutrino $\nu_s$ with Majorana mass 
$m_s$, assumed to be negligibly small for having an observable effect
on the decays of $B$ mesons.
For simplicity we will assume that $\hat y_\nu^{i3} = 0$ so that $\nu_s$ gets
no Dirac mass from the VEV of $H_1$.
After electroweak symmetry breaking, when $H_1$ has obtained the VEV $v/\sqrt{2}$, 
and when the heavy states are integrated out, a Majorana mass matrix 
is generated for the light neutrinos
through the seesaw mechanism,
\be
	m_\nu = \sfrac12\, v^2\, \hat y_\nu^T \hat M_s^{-1} \hat y_\nu
\ee
where $\hat M_s$ is the submatrix of $M_s$ containing the large eigenvalues.
Despite having only two heavy sterile neutrinos, the seesaw mechanism works as usual
to explain the small masses of the active neutrinos.
The neutrino mass matrix $m_\nu$ gets diagonalized by the unitary 
transformation $\nu_L\to L_\nu\nu_i$ where $\nu_i$ denotes the mass
eigenstates.

In the mass eigenbasis for the fermions and scalars, the new Yukawa 
couplings of neutral scalars to neutrinos take the form
\bea
\label{ynu1}
	{\cal L}_{Y_\nu} &=& -{1\over\sqrt{2}}
	\sum_{\phi=h,H,A} y^\nu_{\phi i}\, 
	\bar \nu_i \phi  P_{\sss R} \nu_s +{\rm h.c.} \\
	y^\nu_{h i} &=& \cba\, \rho_\nu^{i}, \nn \\
	y^\nu_{Hi} &=& 
	-\sba\, \rho_\nu^{i}, \nn\\
	y^f_{Ai} &=& i\rho_\nu^{i} \nn
\eea 
where $\rho_\nu = L_\nu^\dagger\, \hat\rho_\nu$ is a vector in the neutrino
flavor space.  The charged Higgs couples to the neutrinos via 
\bea
\label{chargedLnu}
	{\cal L} &=& -H^+\left[\bar\nu\left(U_{\nu}^\dagger\,\rho_e\right)\, P_{\sss R}\, e
	- \bar\nu_s\,(\rho_\nu^\dagger U_\nu)\, P_{\sss L}\,e\right]
\eea
where $U_{\nu} = L_\nu^\dagger \, L_e$ is the PMNS neutrino mixing matrix.
We note that if $\hat y_e$ was 
originally diagonal for some reason, then $\rho_\nu^\dagger U_\nu = 
\hat\rho_\nu$, so that having $\hat\rho_\nu^i \cong \bar\rho_\nu
\delta_{i\tau}$ in the original field basis would explain why
$(\rho_\nu^\dagger U_\nu)^\tau$ is the dominant component of 
$(\rho_\nu^\dagger U_\nu)$ despite the large mixing angles in $U_\nu$.
We will make this assumption in the following, and for simplicity $\rho_e=0$,
so that the charged Higgs coupling to leptons reduces to 
$\bar\rho_\nu H^+\bar\nu_s\tau_L + {\rm h.c.}$.

\subsection{Refitting $R(D)$}

To fit the $R(D)$ observations with this model, we must take into account
that the new amplitude for $B\to D^{(*)}\tau\nu_s$ no longer interferes with
the SM contribution due to the different neutrino flavor.  To make the proper
adjustment it is useful to parametrize the interference effect that occurs
in the original model, where the ratios depend upon $x_\pm = (C^{cb}_{S_R}\pm C^{cb}_{S_L})/C^{cb}_{\sss
SM}$ with $C^{cb}_{\sss SM} = 2\sqrt{2}\,G_F V_{cb}$, as 
\cite{Crivellin:2012ye}
\bea
\label{RDx}
	R(D) &=& R(D)_{\sss SM} (1 +1.5\, x_+ + 1.0\, x_+^2)\nonumber\\
	R(D^*) &=& R(D^*)_{\sss SM} (1 + 0.12\, x_- + 0.05\, x_-^2)
\eea 
The fit (\ref{BDfits2a},\ref{BDfits2b}) of ref.\ \cite{Freytsis:2015qca}
corresponds to $(x_+,\, x_-) = (0.17,\, 1.66)$.  In the sterile neutrino
model, (\ref{RDx}) is modified by omitting the terms linear in $x_\pm$.
We find that $(x_+,\, x_-) \to (0.53,\,2.59)$ to compensate for this change,
leading to the Wilson coefficients (\ref{BDfits1}).

This rescaling ignores the effect of having no interference on the decay
spectra, where the NP contribution to the amplitude multiplies $q^2$,
the invariant lepton pair mass squared \cite{Kamenik:2008tj}. 
Therefore the decay distribution will have a larger $q^4$ contribution
and smaller $q^2$, hardening the spectrum.  We leave to future work
to quantify the effect of this on the fits.  Here it is
mainly important that $C^{cb}_{S_R}$ remains relatively large, which was the
motivation for this study.

Then eq.\ (\ref{BDham}) implies
\bea
\label{BDfitsAa}
	{C^{cb}_{S_R}\over\Lambda^2} &\cong& {\bar\rho_\nu^* 
	(V\rho_d)^{cb}
	\over m_\pm^2}\ \ \,\cong\ \  {2.1\over {\rm TeV}^2}\\
\label{BDfitsAb}
	{C^{cb}_{S_L}\over\Lambda^2} &\cong& -{\bar\rho_\nu^* 
	(\rho_u^\dagger V)^{cb}
	\over m_\pm^2}
	\cong -{1.4\over {\rm TeV}^2}
\eea
Comparison with (\ref{BDfits2a},\ref{BDfits2b}) implies that 
$|\bar\rho_\nu\, \rho_{u,d}|$, must be larger than in 
our previous determination
by a factor of 1.7, while the parameter $\eta$ becomes smaller, $\eta\cong
0.67$.

\begin{figure}[t]
\centerline{
\includegraphics[width=0.7\columnwidth]{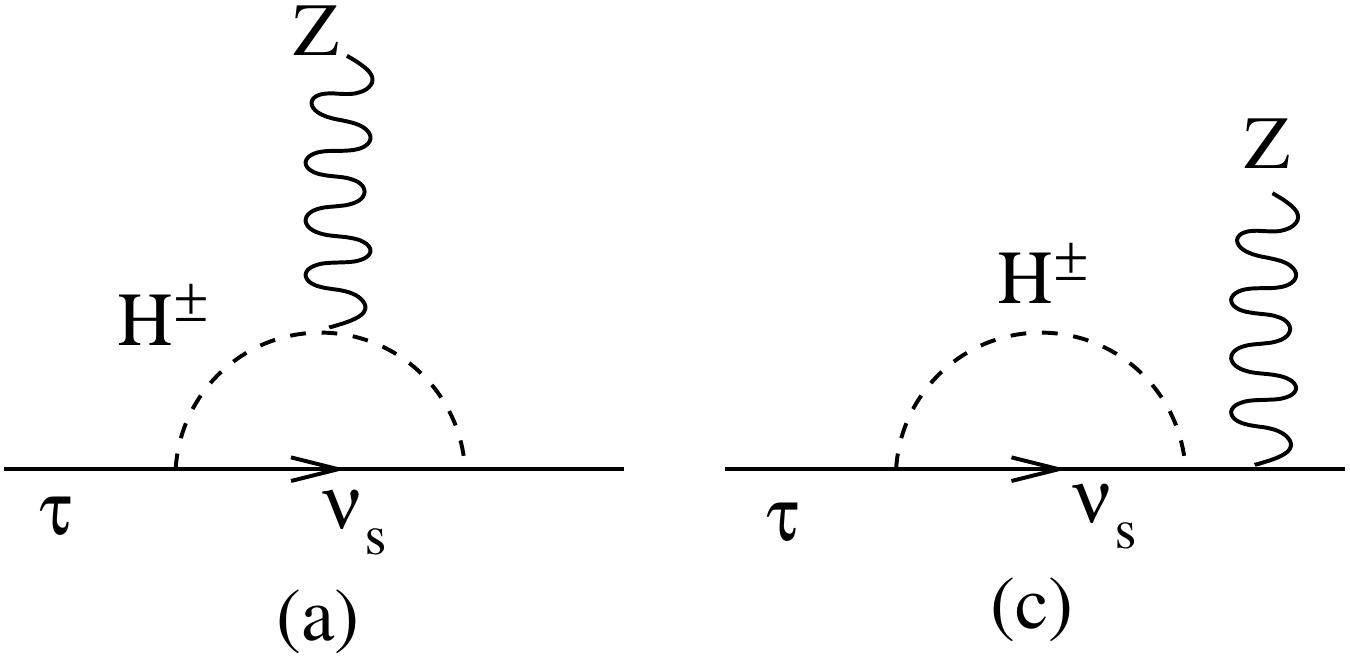}}
\caption{Diagrams contributing to $Z\to\ttpm$ in sterile neutrino
model.  Diagram (b) as in fig.\ \ref{Zdecay} is not present.}
\label{Zdecay2}
\end{figure}

\subsection{$Z\to\ttpm$ and $Z\to\nu_\tau\bar\nu_\tau$}

In contrast to the case of the $\rhoett$ coupling, there are only
two diagrams contributing to $Z\to\ttpm$ with the $\bar\rho_\nu$ coupling,
shown in fig.\ \ref{Zdecay2}.  They involve only exchange of $H^\pm$,
giving the effective interaction term
\bea
\mathcal{L}_{\rm eff,\tau} &=& -(g_{\tau_L} + \delta g_{\tau_L}) 
\,(\bar\tau_L\gamma^\mu\tau_L)\, Z^\mu \nonumber\\
	\hbox{with}\ g_{\tau_L} &=& -\sfrac12 g_Z\,(1-2\sw^2),\quad
	g_Z = {e\over \cw\sw}\nonumber\\
	\delta g_{\tau_L} &=& -{|\bar\rho_\nu|^2 g_Z\over 32\pi^2}\,
F_{\tau_L},
	\nonumber\\
	F_{\tau_L} &=& -\sfrac12(1-2\sw^2)\,(F_a^\pm + \sfrac12 F_c^\pm) 
\label{leff_tau}
\eea
Applying the limit (\ref{deltaR},\ref{dRlimit}), we find 
\be
	|\bar\rho_\nu| < 2.9 \left(m_\pm\over m_Z\right)^2
\ee
which is less restrictive than the analogous bound on $\rhoett$.
For $m_\pm = 100{\rm\,GeV}$, we find $|\bar\rho_\nu| < 3.5$, and in general
the bound is closely numerically fit by $|\bar\rho_\nu| < 2.9\,(m_\pm/m_Z)^2$.
Combining this with (\ref{BDfitsAa}) puts a lower bound on the
$\rho_d$ couplings,  
\be
	{(V\rho_d)^{cb}} > 6.1\times 10^{-3}
\label{Vrhodcb}
\ee

For the new contribution to $Z\to\nu_\tau\bar\nu_\tau$, the relevant
expressions are as in (\ref{leff_nutau}) with the replacements
$\rhoett\to\bar\rho_\nu$ and 
\be
	F_{\nu_\tau} = \sfrac12 F_a^0 + \sfrac14 F_c^0 
\label{fnutau}
\ee
This evaluates to be $0.5-0.6$ times smaller than $F_{\nu_\tau}$ in the
original model, leading to a smaller contribution to the invisible $Z$
width.  In the alternative model, there is also a new contribution
$Z\to\nu_s\nu_s$, but it does not interfere with any SM amplitude so it
is negligible.

\begin{figure}[t]
\centerline{
\includegraphics[width=0.8\columnwidth]{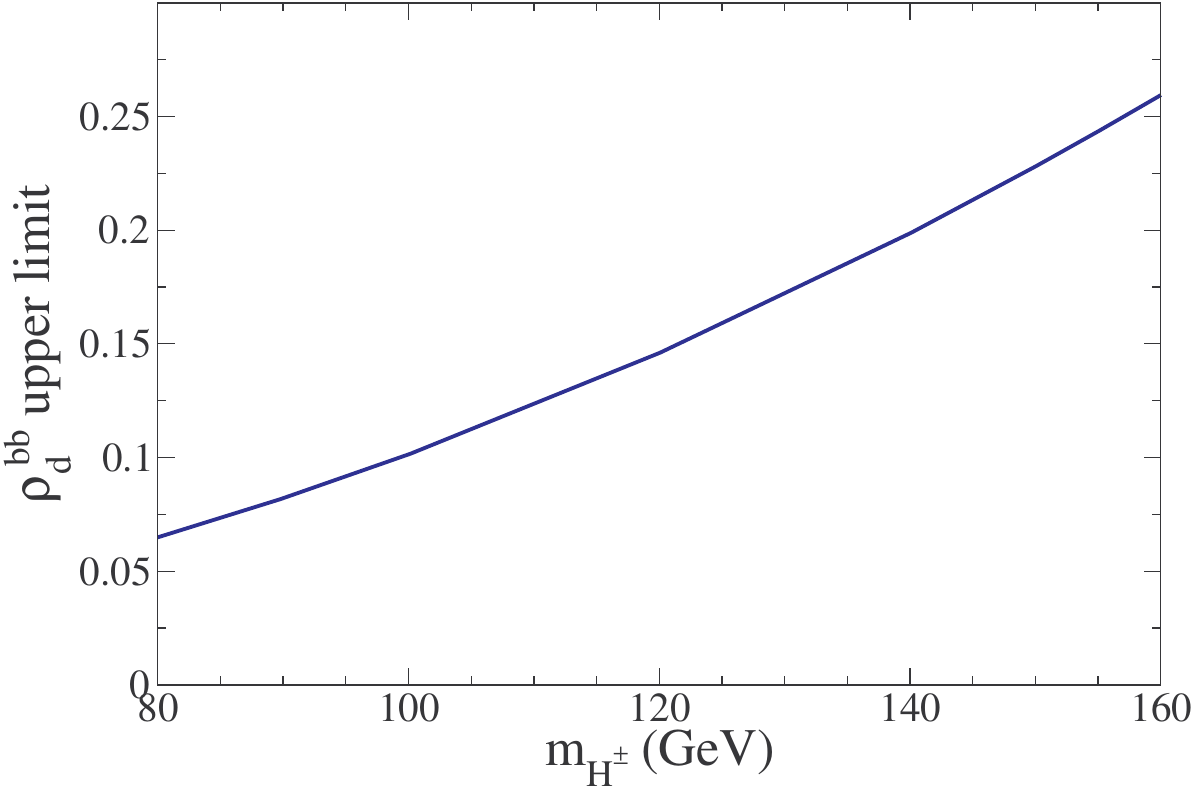}}
\caption{Upper limit on $\rho_d^{bb}$ inferred from CMS search 
\cite{Khachatryan:2015qxa}
for $H^+\to\tau\nu$ as a function of charged Higgs mass $m_\pm$,
for $B(H^\pm\to\tau\nu) = 1$.}
\label{rhodbound}
\end{figure}

\subsection{Collider constraints}
With $H^0$ decaying nearly 100\% to $\nu_s\nu_\tau$, LHC constraints from
neutral Higgs searches are essentially removed.  We are free to take
$m_H\cong 175{\rm\,GeV}$ for example.  Such a value is just compatible
with EWPD constraints (the $\rho$ parameter) on the $m_H$-$m_\pm$ mass
splitting if $m_\pm = 100{\rm\,GeV}$.
Then (\ref{rhodsb_bound}) is 
marginally satisfied
(taking it to now apply to $B_s\to\nu\nu$ decays)
with a value that is compatible with (\ref{Vrhodcb}) even if $\rho_d^{bb}=0$.
In this limiting case we have
\be
	\bar\rho_\nu\,\rho_d^{sb} = 0.021
\label{special_case}
\ee
Then
$\rho_d^{sb}$ is the dominant coupling in $\rho_d$ and does the job
of providing large enough $C^{cb}_{S_R}$ coefficient for fitting $R(D)$. 
With negligible $\rho_d^{bb}$, 
the production of $H^0$ is nullified by any other means than the
subdominant electroweak $Z^*\to HA$ or $W^*\to H^\pm H$ processes.  
Moreover with $\rho_d^{bb}$ very small, the branching ratio $B(t\to H^+b)$
becomes negligible, since $\rho_d^{bb}\to V_{ts}\rho_d^{sb}$ in eq.\
(\ref{BRs}), circumventing the search for charged Higgs bosons. 

The combination $\bar\rho_\nu\cba$ is constrained by the invisible
width of the Higgs boson due to $h\to\nu_s\nu_\tau$, limited to 
$B(h\to\nu_s\nu_\tau) < 36\%$ by CMS \cite{CMS:2015naa}. This implies
$|\bar\rho_\nu| < 1.7\times{10^{-2}/|\cba|}$.  A more stringent bound comes
from the degradation of the total Higgs signal strength 
$\mu \cong 1.1\pm 0.1$ \cite{Higgs_constraints} by invisible decays (not
compensated by any increase in production), 
which implies $\Delta B(h\to\nu\nu) \cong
1 - \mu = -0.1\pm 0.1$ \cite{Espinosa:2012vu}.  The $2\sigma$ upper bound
implies 
\be
	|\bar\rho_\nu| <
	{9\times 10^{-3}\over |\cba|}
\nonumber
\ee
Since (\ref{special_case}) is compatible with both $B_s\to\nu\nu$ and
$R(D)$, we are free to take larger mixing angle and smaller $\bar\rho_\nu$,
for example $\cba=10^{-2}$ and $\bar\rho_\nu = 0.9$, alleviating the mild
naturalness tension for keeping $\cba$ very small, that we encountered in 
the model with $\rhoett$.

We have the freedom to deviate from the limiting case (\ref{special_case})
by turning on $\rho_d^{bb}$ again, such that (\ref{BDfitsAa}) is fulfilled by
a linear combination of $\rho_{d}^{sb}$ and $\rho_{d}^{bb}$.  This reduces
the branching ratio $B(B_s\to\nu\nu)$ and increases that of $t\to H^+ b$
so that CMS searches for $H^\pm\to\tau\nu$ apply.  The resulting
constraint on $\rho_d^{bb}$, plotted in fig.\ \ref{rhodbound},
prevents us from attributing more than 70\% of $(V\rho_d)^{cb}$ (controlling
the Wilson coefficient $C^{cb}_{S_R}$) to the contribution from
$V_{cb}\,\rho_d^{bb}$.  In this other limiting case, the branching ratio for
$B_s\to\nu\nu$ is reduced to the level of $0.5\%$.  Thus while the
alternative version of the model is less constrained, it still predicts
a significant contribution to the invisible width of $B_s$.

\end{appendix}

\end{document}